
%
%
%
\def\unredoffs{} \def\redoffs{\voffset=-.31truein\hoffset=-.59truein}
\def\speclscape{\special{ps: landscape}}
%
%
%
%
\newbox\leftpage \newdimen\fullhsize \newdimen\hstitle \newdimen\hsbody
\tolerance=1000\hfuzz=2pt
\catcode`\@=11 
\def\bigans{b }
\def\answ{b }
\ifx\answ\bigans\message{(This will come out unreduced.}
\magnification=1200\unredoffs\baselineskip=16pt plus 2pt minus 1pt
\hsbody=\hsize \hstitle=\hsize 
\else\message{(This will be reduced.} \let\l@r=L
\magnification=1000\baselineskip=16pt plus 2pt minus 1pt \vsize=7truein
\redoffs \hstitle=8truein\hsbody=4.75truein\fullhsize=10truein\hsize=\hsbody
\output={\ifnum\pageno=0 
  \shipout\vbox{\speclscape{\hsize\fullhsize\makeheadline}
    \hbox to \fullhsize{\hfill\pagebody\hfill}}\advancepageno
  \else
  \almostshipout{\leftline{\vbox{\pagebody\makefootline}}}\advancepageno
  \fi}
\def\almostshipout#1{\if L\l@r \count1=1 \message{[\the\count0.\the\count1]}
      \global\setbox\leftpage=#1 \global\let\l@r=R
 \else \count1=2
  \shipout\vbox{\speclscape{\hsize\fullhsize\makeheadline}
      \hbox to\fullhsize{\box\leftpage\hfil#1}}  \global\let\l@r=L\fi}
\fi
%
\newcount\yearltd\yearltd=\year\advance\yearltd by -1900

\def\Title#1#2{\nopagenumbers\abstractfont\hsize=\hstitle\rightline{#1}%
\vskip 1in\centerline{\titlefont #2}\abstractfont\vskip .5in\pageno=0}
\def\Date#1{\vfill\leftline{#1}\tenpoint\supereject\global\hsize=\hsbody%
\footline={\hss\tenrm\folio\hss}}
%

\def\draftmode{\message{ DRAFTMODE }\def\draftdate{{\rm preliminary draft:
\number\month/\number\day/\number\yearltd\ \ \hourmin}}%
\headline={\hfil\draftdate}\writelabels\baselineskip=20pt plus 2pt minus 2pt
 {\count255=\time\divide\count255 by 60 \xdef\hourmin{\number\count255}
  \multiply\count255 by-60\advance\count255 by\time
  \xdef\hourmin{\hourmin:\ifnum\count255<10 0\fi\the\count255}}}
\def\nolabels{\def\wrlabeL##1{}\def\eqlabeL##1{}\def\reflabeL##1{}}
\def\writelabels{\def\wrlabeL##1{\leavevmode\vadjust{\rlap{\smash%
{\line{{\escapechar=` \hfill\rlap{\sevenrm\hskip.03in\string##1}}}}}}}%
\def\eqlabeL##1{{\escapechar-1\rlap{\sevenrm\hskip.05in\string##1}}}%
\def\reflabeL##1{\noexpand\llap{\noexpand\sevenrm\string\string\string##1}}}
\nolabels
%
\global\newcount\secno \global\secno=0
\global\newcount\meqno \global\meqno=1
\def\newsec#1{\global\advance\secno by1\message{(\the\secno. #1)}
\global\subsecno=0\eqnres@t\noindent{\bf\the\secno. #1}
\writetoca{{\secsym} {#1}}\par\nobreak\medskip\nobreak}
\def\eqnres@t{\xdef\secsym{\the\secno.}\global\meqno=1\bigbreak\bigskip}
\def\sequentialequations{\def\eqnres@t{\bigbreak}}\xdef\secsym{}
\global\newcount\subsecno \global\subsecno=0
\def\subsec#1{\global\advance\subsecno by1\message{(\secsym\the\subsecno.
#1)}
\ifnum\lastpenalty>9000\else\bigbreak\fi
\noindent{\it\secsym\the\subsecno. #1}\writetoca{\string\quad
{\secsym\the\subsecno.} {#1}}\par\nobreak\medskip\nobreak}
\def\appendix#1#2{\global\meqno=1\global\subsecno=0\xdef\secsym{\hbox{#1.}}
\bigbreak\bigskip\noindent{\bf Appendix #1. #2}\message{(#1. #2)}
\writetoca{Appendix {#1.} {#2}}\par\nobreak\medskip\nobreak}
%
%
\def\eqnn#1{\xdef #1{(\secsym\the\meqno)}\writedef{#1\leftbracket#1}%
\global\advance\meqno by1\wrlabeL#1}
\def\eqna#1{\xdef #1##1{\hbox{$(\secsym\the\meqno##1)$}}
\writedef{#1\numbersign1\leftbracket#1{\numbersign1}}%
\global\advance\meqno by1\wrlabeL{#1$\{\}$}}
\def\eqn#1#2{\xdef #1{(\secsym\the\meqno)}\writedef{#1\leftbracket#1}%
\global\advance\meqno by1$$#2\eqno#1\eqlabeL#1$$}
%
\newskip\footskip\footskip14pt plus 1pt minus 1pt 
\def\footnotefont{\ninepoint}\def\f@t#1{\footnotefont #1\@foot}
\def\f@@t{\baselineskip\footskip\bgroup\footnotefont\aftergroup\@foot\let\next}
\setbox\strutbox=\hbox{\vrule height9.5pt depth4.5pt width0pt}
\global\newcount\ftno \global\ftno=0
\def\foot{\global\advance\ftno by1\footnote{$^{\the\ftno}$}}
%
\newwrite\ftfile
\def\footend{\def\foot{\global\advance\ftno by1\chardef\wfile=\ftfile
$^{\the\ftno}$\ifnum\ftno=1\immediate\openout\ftfile=foots.tmp\fi%
\immediate\write\ftfile{\noexpand\smallskip%
\noexpand\item{f\the\ftno:\ }\pctsign}\findarg}%
\def\footatend{\vfill\eject\immediate\closeout\ftfile{\parindent=20pt
\centerline{\bf Footnotes}\nobreak\bigskip\input foots.tmp }}}
\def\footatend{}
%
%
\global\newcount\refno \global\refno=1
\newwrite\rfile
\def\ref{[\the\refno]\nref}
\def\nref#1{\xdef#1{[\the\refno]}\writedef{#1\leftbracket#1}%
\ifnum\refno=1\immediate\openout\rfile=refs.tmp\fi
\global\advance\refno by1\chardef\wfile=\rfile\immediate
\write\rfile{\noexpand\item{#1\ }\reflabeL{#1\hskip.31in}\pctsign}\findarg}
\def\findarg#1#{\begingroup\obeylines\newlinechar=`\^^M\pass@rg}
{\obeylines\gdef\pass@rg#1{\writ@line\relax #1^^M\hbox{}^^M}%
\gdef\writ@line#1^^M{\expandafter\toks0\expandafter{\striprel@x #1}%
\edef\next{\the\toks0}\ifx\next\em@rk\let\next=\endgroup\else\ifx\next\empty%
\else\immediate\write\wfile{\the\toks0}\fi\let\next=\writ@line\fi\next\relax}}
\def\striprel@x#1{} \def\em@rk{\hbox{}}
\def\lref{\begingroup\obeylines\lr@f}
\def\lr@f#1#2{\gdef#1{\ref#1{#2}}\endgroup\unskip}
\def\semi{;\hfil\break}
\def\addref#1{\immediate\write\rfile{\noexpand\item{}#1}} 
\def\startrefs#1{\immediate\openout\rfile=refs.tmp\refno=#1}
\def\xref{\expandafter\xr@f}\def\xr@f[#1]{#1}
\def\refs#1{\count255=1[\r@fs #1{\hbox{}}]}
\def\r@fs#1{\ifx\und@fined#1\message{reflabel \string#1 is undefined.}%
\nref#1{need to supply reference \string#1.}\fi%
\vphantom{\hphantom{#1}}\edef\next{#1}\ifx\next\em@rk\def\next{}%
\else\ifx\next#1\ifodd\count255\relax\xref#1\count255=0\fi%
\else#1\count255=1\fi\let\next=\r@fs\fi\next}
%

%
\newwrite\ffile\global\newcount\figno \global\figno=1
\def\fig{fig.~\the\figno\nfig}
\def\nfig#1{\xdef#1{fig.~\the\figno}%
\writedef{#1\leftbracket fig.\noexpand~\the\figno}%
\ifnum\figno=1\immediate\openout\ffile=figs.tmp\fi\chardef\wfile=\ffile%
\immediate\write\ffile{\noexpand\medskip\noexpand\item{Fig.\ \the\figno. }
\reflabeL{#1\hskip.55in}\pctsign}\global\advance\figno by1\findarg}
\def\xfig{\expandafter\xf@g}\def\xf@g fig.\penalty\@M\ {}
\def\figs#1{figs.~\f@gs #1{\hbox{}}}
\def\f@gs#1{\edef\next{#1}\ifx\next\em@rk\def\next{}\else
\ifx\next#1\xfig #1\else#1\fi\let\next=\f@gs\fi\next}
\newwrite\lfile
{\escapechar-1\xdef\pctsign{\string\%}\xdef\leftbracket{\string\{}
\xdef\rightbracket{\string\}}\xdef\numbersign{\string\#}}

\def\writestop{\def\writestoppt{\immediate\write\lfile{\string\pageno%
\the\pageno\string\startrefs\leftbracket\the\refno\rightbracket%
\string\def\string\secsym\leftbracket\secsym\rightbracket%
\string\secno\the\secno\string\meqno\the\meqno}\immediate\closeout\lfile}}
\def\writestoppt{}\def\writedef#1{}
\def\seclab#1{\xdef #1{\the\secno}\writedef{#1\leftbracket#1}\wrlabeL{#1=#1}}
\def\subseclab#1{\xdef #1{\secsym\the\subsecno}%
\writedef{#1\leftbracket#1}\wrlabeL{#1=#1}}
\newwrite\tfile \def\writetoca#1{}
\def\leaderfill{\leaders\hbox to 1em{\hss.\hss}\hfill}
\def\writetoc{\immediate\openout\tfile=toc.tmp
   \def\writetoca##1{{\edef\next{\write\tfile{\noindent ##1
   \string\leaderfill {\noexpand\number\pageno} \par}}\next}}}

%
\catcode`\@=12 
%
\edef\tfontsize{\ifx\answ\bigans scaled\magstep3\else scaled\magstep4\fi}
\font\titlerm=cmr10 \tfontsize \font\titlerms=cmr7 \tfontsize
\font\titlermss=cmr5 \tfontsize \font\titlei=cmmi10 \tfontsize
\font\titleis=cmmi7 \tfontsize \font\titleiss=cmmi5 \tfontsize
\font\titlesy=cmsy10 \tfontsize \font\titlesys=cmsy7 \tfontsize
\font\titlesyss=cmsy5 \tfontsize \font\titleit=cmti10 \tfontsize
\skewchar\titlei='177 \skewchar\titleis='177 \skewchar\titleiss='177
\skewchar\titlesy='60 \skewchar\titlesys='60 \skewchar\titlesyss='60
\def\titlefont{\def\rm{\fam0\titlerm}
\textfont0=\titlerm \scriptfont0=\titlerms \scriptscriptfont0=\titlermss
\textfont1=\titlei \scriptfont1=\titleis \scriptscriptfont1=\titleiss
\textfont2=\titlesy \scriptfont2=\titlesys \scriptscriptfont2=\titlesyss
\textfont\itfam=\titleit \def\it{\fam\itfam\titleit}\rm}
 \ifx\answ\bigans\else scaled\magstep1\fi
\ifx\answ\bigans\def\abstractfont{\tenpoint}\else
\font\abssl=cmsl10 scaled \magstep1
\font\absrm=cmr10 scaled\magstep1 \font\absrms=cmr7 scaled\magstep1
\font\absrmss=cmr5 scaled\magstep1 \font\absi=cmmi10 scaled\magstep1
\font\absis=cmmi7 scaled\magstep1 \font\absiss=cmmi5 scaled\magstep1
\font\abssy=cmsy10 scaled\magstep1 \font\abssys=cmsy7 scaled\magstep1
\font\abssyss=cmsy5 scaled\magstep1 \font\absbf=cmbx10 scaled\magstep1
\skewchar\absi='177 \skewchar\absis='177 \skewchar\absiss='177
\skewchar\abssy='60 \skewchar\abssys='60 \skewchar\abssyss='60
\def\abstractfont{\def\rm{\fam0\absrm}
\textfont0=\absrm \scriptfont0=\absrms \scriptscriptfont0=\absrmss
\textfont1=\absi \scriptfont1=\absis \scriptscriptfont1=\absiss
\textfont2=\abssy \scriptfont2=\abssys \scriptscriptfont2=\abssyss
\textfont\itfam=\bigit \def\it{\fam\itfam\bigit}\def\footnotefont{\tenpoint}%
\textfont\slfam=\abssl \def\sl{\fam\slfam\abssl}%
\textfont\bffam=\absbf \def\bf{\fam\bffam\absbf}\rm}\fi
\def\tenpoint{\def\rm{\fam0\tenrm}
\textfont0=\tenrm \scriptfont0=\sevenrm \scriptscriptfont0=\fiverm
\textfont1=\teni  \scriptfont1=\seveni  \scriptscriptfont1=\fivei
\textfont2=\tensy \scriptfont2=\sevensy \scriptscriptfont2=\fivesy
\textfont\itfam=\tenit
\def\it{\fam\itfam\tenit}\def\footnotefont{\ninepoint}%
\textfont\bffam=\tenbf \def\bf{\fam\bffam\tenbf}\def\sl{\fam\slfam\tensl}\rm}
\font\ninerm=cmr9 \font\sixrm=cmr6 \font\ninei=cmmi9 \font\sixi=cmmi6
\font\ninesy=cmsy9 \font\sixsy=cmsy6 \font\ninebf=cmbx9
\font\nineit=cmti9 \font\ninesl=cmsl9 \skewchar\ninei='177
\skewchar\sixi='177 \skewchar\ninesy='60 \skewchar\sixsy='60
\def\ninepoint{\def\rm{\fam0\ninerm}
\textfont0=\ninerm \scriptfont0=\sixrm \scriptscriptfont0=\fiverm
\textfont1=\ninei \scriptfont1=\sixi \scriptscriptfont1=\fivei
\textfont2=\ninesy \scriptfont2=\sixsy \scriptscriptfont2=\fivesy
\textfont\itfam=\ninei \def\it{\fam\itfam\nineit}\def\sl{\fam\slfam\ninesl}%
\textfont\bffam=\ninebf \def\bf{\fam\bffam\ninebf}\rm}
%
%

\hyphenation{anom-aly anom-alies coun-ter-term coun-ter-terms}
\def\inv{^{\raise.15ex\hbox{${\scriptscriptstyle -}$}\kern-.05em 1}}

\def\Dsl{\,\raise.15ex\hbox{/}\mkern-13.5mu D} 
\def\dsl{\raise.15ex\hbox{/}\kern-.57em\partial}

\def\tr{{\rm tr}} \def\Tr{{\rm Tr}}
\font\bigit=cmti10 scaled \magstep1
\def\lspace{\ifx\answ\bigans{}\else\qquad\fi}
\def\lbspace{\ifx\answ\bigans{}\else\hskip-.2in\fi} 
\def\boxeqn#1{\vcenter{\vbox{\hrule\hbox{\vrule\kern3pt\vbox{\kern3pt
           \hbox{${\displaystyle #1}$}\kern3pt}\kern3pt\vrule}\hrule}}}
\def\mbox#1#2{\vcenter{\hrule \hbox{\vrule height#2in
               \kern#1in \vrule} \hrule}}  
%
   
\def\CL{{\cal L}}   
   
\def\e#1{{\rm e}^{^{\textstyle#1}}}

\def\darr#1{\raise1.5ex\hbox{$\leftrightarrow$}\mkern-16.5mu #1}

\def\roughly#1{\raise.3ex\hbox{$#1$\kern-.75em\lower1ex\hbox{$\sim$}}}



\def\IB{\relax\hbox{$\inbar\kern-.3em{\rm B}$}}
\def\IC{\relax\hbox{$\inbar\kern-.3em{\rm C}$}}
\def\ID{\relax\hbox{$\inbar\kern-.3em{\rm D}$}}
\def\IE{\relax\hbox{$\inbar\kern-.3em{\rm E}$}}
\def\IF{\relax\hbox{$\inbar\kern-.3em{\rm F}$}}
\def\IG{\relax\hbox{$\inbar\kern-.3em{\rm G}$}}
\def\IGa{\relax\hbox{${\rm I}\kern-.18em\Gamma$}}
\def\IH{\relax{\rm I\kern-.18em H}}
\def\IK{\relax{\rm I\kern-.18em K}}
\def\II{\relax{\rm I\kern-.18em I}}
\def\IL{\relax{\rm I\kern-.18em L}}
\def\IP{\relax{\rm I\kern-.18em P}}
\def\IR{\relax{\rm I\kern-.18em R}}
\def\IZ{\relax\ifmmode\mathchoice {\hbox{\cmss Z\kern-.4em Z}}{\hbox{\cmss
Z\kern-.4em Z}} {\lower.9pt\hbox{\cmsss Z\kern-.4em Z}}
{\lower1.2pt\hbox{\cmsss Z\kern-.4em Z}}\else{\cmss Z\kern-.4em Z}\fi}

\def\IB{\relax{\rm I\kern-.18em B}}
\def\IC{{\relax\hbox{$\inbar\kern-.3em{\rm C}$}}}
\def\ID{\relax{\rm I\kern-.18em D}}
\def\IE{\relax{\rm I\kern-.18em E}}
\def\IF{\relax{\rm I\kern-.18em F}}


\def\CL {{\cal L}}

\def\CW {{\cal W}}

\def\p{\partial}
\def\pa{\partial}



\def\cb{{\bar c}}


\def\Tr{{\rm Tr}}


\def\demi{{1\over 2}}

\def\c{\cdot}


\def\f{\phi}    
\def\P{\Psi}    
\def\F{\Phi}

\def\a{\alpha}
\def\b{\beta}
\def\g{\gamma}  
\def\d{\delta}  \def\D{\Delta}
\def\m{\mu}
\def\n{\nu}
\def\r{\rho}
\def\l{\lambda} \def\L{\Lambda}

\def\e{\epsilon}

\def\|{\Big|}
\def\({\Big(}   \def\){\Big)}
\def\[{\Big[}   \def\]{\Big]}



\def\paper#1#2#3#4{#1, {\sl #2}, #3 {\tt #4}}

\def\hh{hep-th/}


\def\PLB#1#2#3{Phys. Lett.~{\bf B#1} (#2) #3}
\def\NPB#1#2#3{Nucl. Phys.~{\bf B#1} (#2) #3}
\def\PRL#1#2#3{Phys. Rev. Lett.~{\bf #1} (#2) #3}
\def\CMP#1#2#3{Comm. Math. Phys.~{\bf #1} (#2) #3}
\def\PRD#1#2#3{Phys. Rev.~{\bf D#1} (#2) #3}
\def\MPL#1#2#3{Mod. Phys. Lett.~{\bf #1} (#2) #3}
\def\IJMP#1#2#3{Int. Jour. Mod. Phys.~{\bf #1} (#2) #3}


\def\unlockat{\catcode`\@=11}
\def\lockat{\catcode`\@=12}

\unlockat


\def\newsec#1{\global\advance\secno by1\message{(\the\secno. #1)}
\global\subsecno=0\global\subsubsecno=0\eqnres@t\noindent {\bf\the\secno. #1}
\writetoca{{\secsym} {#1}}\par\nobreak\medskip\nobreak}
\global\newcount\subsecno \global\subsecno=0
\def\subsec#1{\global\advance\subsecno by1\message{(\secsym\the\subsecno.
#1)}
\ifnum\lastpenalty>9000\else\bigbreak\fi\global\subsubsecno=0
\noindent{\it\secsym\the\subsecno. #1}
\writetoca{\string\quad {\secsym\the\subsecno.} {#1}}
\par\nobreak\medskip\nobreak}
\global\newcount\subsubsecno \global\subsubsecno=0
\def\subsubsec#1{\global\advance\subsubsecno by1
\message{(\secsym\the\subsecno.\the\subsubsecno. #1)}
\ifnum\lastpenalty>9000\else\bigbreak\fi
\noindent\quad{\secsym\the\subsecno.\the\subsubsecno.}{#1}
\writetoca{\string\qquad{\secsym\the\subsecno.\the\subsubsecno.}{#1}}
\par\nobreak\medskip\nobreak}

\def\subsubseclab#1{\DefWarn#1\xdef #1{\noexpand\hyperref{}{subsubsection}%
{\secsym\the\subsecno.\the\subsubsecno}%
{\secsym\the\subsecno.\the\subsubsecno}}%
\writedef{#1\leftbracket#1}\wrlabeL{#1=#1}}
\lockat

\def\dbend{\lower3.5pt\hbox{\manual\char127}}


\def\boxit#1{\vbox{\hrule\hbox{\vrule\kern8pt
\vbox{\hbox{\kern8pt}\hbox{\vbox{#1}}\hbox{\kern8pt}}
\kern8pt\vrule}\hrule}}

\def\mathboxit#1{\vbox{\hrule\hbox{\vrule\kern8pt\vbox{\kern8pt
\hbox{$\displaystyle #1$}\kern8pt}\kern8pt\vrule}\hrule}}


\def\inbar{\,\vrule height1.5ex width.4pt depth0pt}

\font\cmss=cmss10 \font\cmsss=cmss10 at 7pt


\lref\simons{ J. Cheeger and J. Simons, {\it Differential Characters and
Geometric Invariants},  Stony Brook Preprint, (1973), unpublished.}

\lref\cargese{ L.~Baulieu, {\it Algebraic quantization of gauge theories},
Perspectives in fields and particles, Plenum Press, eds. Basdevant-Levy,
Cargese Lectures 1983}

\lref\antifields{ L. Baulieu, M. Bellon, S. Ouvry, C.Wallet, Phys.Letters
B252 (1990) 387; M.  Bocchichio, Phys. Lett. B187 (1987) 322;  Phys. Lett. B
192 (1987) 31; R.  Thorn    Nucl. Phys.   B257 (1987) 61. }

\lref\thompson{ George Thompson,  Annals Phys. 205 (1991) 130; J.M.F.
Labastida, M. Pernici, Phys. Lett. 212B  (1988) 56; D. Birmingham, M.Blau,
M. Rakowski and G.Thompson, Phys. Rept. 209 (1991) 129.}

\lref\tonin{ Tonin}

\lref\wittensix{ E.  Witten, {\it New  Gauge  Theories In Six Dimensions},
\hh{9710065}. }

\lref\orlando{ O. Alvarez, L. A. Ferreira and J. Sanchez Guillen, {\it  A New
Approach to Integrable Theories in any Dimension}, hep-th/9710147.}

\lref\wittentopo{ E.  Witten,  {\it  Topological Quantum Field Theory},
\hh9403195, Commun.  Math. Phys.  {117} (1988)353.  }

\lref\wittentwist{ E.  Witten, {\it Supersymmetric Yang--Mills theory on a
four-manifold}, J.  Math.  Phys.  {35} (1994) 5101.}

\lref\west{ L.~Baulieu, P.~West, {\it Six Dimensional TQFTs and  Self-dual
Two-Forms,} Phys.Lett. B {\bf 436 } (1998) 97, /hep-th/9805200}

\lref\bv{ I.A. Batalin and V.A. Vilkowisky,    Phys. Rev.   D28  (1983)
2567\semi M. Henneaux,  Phys. Rep.  126   (1985) 1\semi M. Henneaux and C.
Teitelboim, {\it Quantization of Gauge Systems}
  Princeton University Press,  Princeton (1992).}

\lref\kyoto{ L. Baulieu, E. Bergschoeff and E. Sezgin, Nucl. Phys.
B307(1988)348\semi L. Baulieu,   {\it Field Antifield Duality, p-Form Gauge
Fields
   and Topological Quantum Field Theories}, hep-th/9512026,
   Nucl. Phys. B478 (1996) 431.  }

\lref\sourlas{ G. Parisi and N. Sourlas, {\it Random Magnetic Fields,
Supersymmetry and Negative Dimensions}, Phys. Rev. Lett.  43 (1979) 744;
Nucl.  Phys.  B206 (1982) 321.  }

\lref\SalamSezgin{ A.  Salam  and  E.  Sezgin, {\it Supergravities in
diverse dimensions}, vol.  1, p. 119\semi P.  Howe, G.  Sierra and P.
Townsend, Nucl Phys B221 (1983) 331.}

\lref\nekrasov{ A. Losev, G. Moore, N. Nekrasov, S. Shatashvili, {\it
Four-Dimensional Avatars of Two-Dimensional RCFT},  hep-th/9509151, Nucl.
Phys.  Proc.  Suppl.   46 (1996) 130\semi L.  Baulieu, A.  Losev,
N.~Nekrasov  {\it Chern-Simons and Twisted Supersymmetry in Higher
Dimensions},  hep-th/9707174, to appear in Nucl.  Phys.  B.  }

\lref\WitDonagi{R.~ Donagi, E.~ Witten, ``Supersymmetric Yang--Mills Theory
and Integrable Systems'', hep-th/9510101, Nucl. Phys.{\bf B}460 (1996)
299-334}
\lref\Witfeb{E.~ Witten, ``Supersymmetric Yang--Mills Theory On A
Four-Manifold,''  hep-th/9403195; J. Math. Phys. {\bf 35} (1994) 5101.}
\lref\Witgrav{E.~ Witten, ``Topological Gravity'', Phys.Lett.206B:601, 1988}
\lref\witaffl{I. ~ Affleck, J.A.~ Harvey and E.~ Witten,
        ``Instantons and (Super)Symmetry Breaking
        in $2+1$ Dimensions'', Nucl. Phys. {\bf B}206 (1982) 413}
\lref\wittabl{E.~ Witten,  ``On $S$-Duality in Abelian Gauge Theory,''
hep-th/9505186; Selecta Mathematica {\bf 1} (1995) 383}
\lref\wittgr{E.~ Witten, ``The Verlinde Algebra And The Cohomology Of The
Grassmannian'',  hep-th/9312104}
\lref\wittenwzw{E. Witten, ``Non abelian bosonization in two dimensions,''
Commun. Math. Phys. {\bf 92} (1984)455 }
\lref\witgrsm{E. Witten, ``Quantum field theory, grassmannians and algebraic
curves,'' Commun.Math.Phys.113:529,1988}
\lref\wittjones{E. Witten, ``Quantum field theory and the Jones
polynomial,'' Commun.  Math. Phys., 121 (1989) 351. }
\lref\witttft{E.~ Witten, ``Topological Quantum Field Theory", Commun. Math.
Phys. {\bf 117} (1988) 353.}
\lref\wittmon{E.~ Witten, ``Monopoles and Four-Manifolds'', hep-th/9411102}
\lref\Witdgt{ E.~ Witten, ``On Quantum gauge theories in two dimensions,''
Commun. Math. Phys. {\bf  141}  (1991) 153}
\lref\witrevis{E.~ Witten,
 ``Two-dimensional gauge theories revisited'', hep-th/9204083; J. Geom.
Phys. 9 (1992) 303-368}
\lref\Witgenus{E.~ Witten, ``Elliptic Genera and Quantum Field Theory'',
Comm. Math. Phys. 109(1987) 525. }
\lref\OldZT{E. Witten, ``New Issues in Manifolds of SU(3) Holonomy,'' {\it
Nucl. Phys.} {\bf B268} (1986) 79 \semi J. Distler and B. Greene, ``Aspects
of (2,0) String Compactifications,'' {\it Nucl. Phys.} {\bf B304} (1988) 1
\semi B. Greene, ``Superconformal Compactifications in Weighted Projective
Space,'' {\it Comm. Math. Phys.} {\bf 130} (1990) 335.}
\lref\bagger{E.~ Witten and J. Bagger, Phys. Lett. {\bf 115B}(1982) 202}
\lref\witcurrent{E.~ Witten,``Global Aspects of Current Algebra'',
Nucl.Phys.B223 (1983) 422\semi ``Current Algebra, Baryons and Quark
Confinement'', Nucl.Phys. B223 (1993) 433}
\lref\Wittreiman{S.B. Treiman, E. Witten, R. Jackiw, B. Zumino, ``Current
Algebra and Anomalies'', Singapore, Singapore: World Scientific ( 1985) }
\lref\Witgravanom{L. Alvarez-Gaume, E.~ Witten, ``Gravitational Anomalies'',
Nucl.Phys.B234:269,1984. }

\lref\nicolai{\paper {H.~Nicolai}{New Linear Systems for 2D Poincar\'e
Supergravities}{\NPB{414}{1994}{299},}{\hh 9309052}.}



\lref\baex{\paper {L.~Baulieu, B.~Grossman}{Monopoles and Topological Field
Theory}{\PLB{214}{1988}{223}.}{}\paper {L.~Baulieu}{Chern-Simons
Three-Dimensional and
Yang--Mills-Higgs Two-Dimensional Systems as Four-Dimensional Topological
Quantum Field Theories}{\PLB{232}{1989}{473}.}{}}

\lref\bg{\paper {L.~Baulieu, B.~Grossman}{Monopoles and Topological Field
Theory}{\PLB{214}{1988}{223}.}{}}

\lref\seibergsix{\paper {N.~Seiberg}{Non-trivial Fixed Points of The
Renormalization Group in Six
 Dimensions}{\PLB{390}{1997}{169}}{\hh 9609161}\semi
\paper {O.J.~Ganor, D.R.~Morrison, N.~Seiberg}{
  Branes, Calabi-Yau Spaces, and Toroidal Compactification of the N=1
  Six-Dimensional $E_8$ Theory}{\NPB{487}{1997}{93}}{\hh 9610251}\semi
\paper {O.~Aharony, M.~Berkooz, N.~Seiberg}{Light-Cone
  Description of (2,0) Superconformal Theories in Six
  Dimensions}{Adv. Theor. Math. Phys. {\bf 2} (1998) 119}{\hh 9712117.}}

\lref\cs{\paper {L.~Baulieu}{Chern-Simons Three-Dimensional and
Yang--Mills-Higgs Two-Dimensional Systems as Four-Dimensional Topological
Quantum Field Theories}{\PLB{232}{1989}{473}.}{}}

\lref\beltrami{\paper {L.~Baulieu, M.~Bellon}{Beltrami Parametrization and
String Theory}{\PLB{196}{1987}{142}}{}\semi
\paper {L.~Baulieu, M.~Bellon, R.~Grimm}{Beltrami Parametrization For
Superstrings}{\PLB{198}{1987}{343}}{}\semi
\paper {R.~Grimm}{Left-Right Decomposition of Two-Dimensional Superspace
Geometry and Its BRS Structure}{Annals Phys. {\bf 200} (1990) 49.}{}}

\lref\bbg{\paper {L.~Baulieu, M.~Bellon, R.~Grimm}{Left-Right Asymmetric
Conformal Anomalies}{\PLB{228}{1989}{325}.}{}}

\lref\bonora{\paper {G.~Bonelli, L.~Bonora, F.~Nesti}{String Interactions
from Matrix String Theory}{\NPB{538}{1999}{100},}{\hh 9807232}\semi
\paper {G.~Bonelli, L.~Bonora, F.~Nesti, A.~Tomasiello}{Matrix String Theory
and its Moduli Space}{}{\hh 9901093.}}

\lref\corrigan{\paper {E.~Corrigan, C.~Devchand, D.B.~Fairlie,
J.~Nuyts}{First Order Equations for Gauge Fields in Spaces of Dimension
Greater Than Four}{\NPB{214}{452}{1983}.}{}}

\lref\acha{\paper {B.S.~Acharya, M.~O'Loughlin, B.~Spence}{Higher
Dimensional Analogues of Donaldson-Witten Theory}{\NPB{503}{1997}{657},}{\hh
9705138}\semi
\paper {B.S.~Acharya, J.M.~Figueroa-O'Farrill, M.~O'Loughlin,
B.~Spence}{Euclidean
  D-branes and Higher-Dimensional Gauge   Theory}{\NPB{514}{1998}{583},}{\hh
  9707118.}}

\lref\Witr{\paper{E.~Witten}{Introduction to Cohomological Field   Theories}
{Lectures at Workshop on Topological Methods in Physics (Trieste, Italy, Jun
11-25, 1990), \IJMP{A6}{1991}{2775}.}{}}

\lref\ohta{\paper {L.~Baulieu, N.~Ohta}{Worldsheets with Extended
Supersymmetry} {\PLB{391}{1997}{295},}{\hh 9609207}.}

\lref\gravity{\paper {L.~Baulieu}{Transmutation of Pure 2-D Supergravity
Into Topological 2-D Gravity and Other Conformal Theories}
{\PLB{288}{1992}{59},}{\hh 9206019.}}

\lref\wgravity{\paper {L.~Baulieu, M.~Bellon, R.~Grimm}{Some Remarks on  the
Gauging of the Virasoro and   $w_{1+\infty}$
Algebras}{\PLB{260}{1991}{63}.}{}}

\lref\fourd{\paper {E.~Witten}{Topological Quantum Field
Theory}{\CMP{117}{1988}{353}}{}\semi
\paper {L.~Baulieu, I.M.~Singer}{Topological Yang--Mills Symmetry}{Nucl.
Phys. Proc. Suppl. {\bf 15B} (1988) 12.}{}}

\lref\topo{\paper {L.~Baulieu}{On the Symmetries of Topological Quantum Field
Theories}{\IJMP{A10}{1995}{4483},}{\hh 9504015}\semi
\paper {R.~Dijkgraaf, G.~Moore}{Balanced Topological Field
Theories}{\CMP{185}{1997}{411},}{\hh 9608169.}}

\lref\wwgravity{\paper {I.~Bakas} {The Large $N$ Limit   of Extended
Conformal Symmetries}{\PLB{228}{1989}{57}.}{}}

\lref\wwwgravity{\paper {C.M.~Hull}{Lectures on $\CW$-Gravity,
$\CW$-Geometry and
$\CW$-Strings}{}{\hh 9302110}, and~references therein.}

\lref\wvgravity{\paper {A.~Bilal, V.~Fock, I.~Kogan}{On the origin of
$W$-algebras}{\NPB{359}{1991}{635}.}{}}

\lref\surprises{\paper {E.~Witten} {Surprises with Topological Field
Theories} {Lectures given at ``Strings 90'', Texas A\&M, 1990,}{Preprint
IASSNS-HEP-90/37.}}

\lref\stringsone{\paper {L.~Baulieu, M.B.~Green, E.~Rabinovici}{A Unifying
Topological Action for Heterotic and  Type II Superstring  Theories}
{\PLB{386}{1996}{91},}{\hh 9606080.}}

\lref\stringsN{\paper {L.~Baulieu, M.B.~Green, E.~Rabinovici}{Superstrings
from   Theories with $N>1$ World Sheet Supersymmetry}
{\NPB{498}{1997}{119},}{\hh 9611136.}}

\lref\bks{\paper {L.~Baulieu, H.~Kanno, I.~Singer}{Special Quantum Field
Theories in Eight and Other Dimensions}{\CMP{194}{1998}{149},}{\hh
9704167}\semi
\paper {L.~Baulieu, H.~Kanno, I.~Singer}{Cohomological Yang--Mills Theory
  in Eight Dimensions}{ Talk given at APCTP Winter School on Dualities in
String Theory (Sokcho, Korea, February 24-28, 1997),} {\hh 9705127.}}

\lref\witdyn{\paper {P.~Townsend}{The eleven dimensional supermembrane
revisited}{\PLB{350}{1995}{184},}{\hh9501068}\semi
\paper{E.~Witten}{String Theory Dynamics in Various Dimensions}
{\NPB{443}{1995}{85},}{\hh 9503124}.}

\lref\bfss{\paper {T.~Banks, W.Fischler, S.H.~Shenker,
L.~Susskind}{$M$-Theory as a Matrix Model~:
A~Conjecture}{\PRD{55}{1997}{5112},} {\hh9610043.}}

\lref\seiberg{\paper {N.~Seiberg}{Why is the Matrix Model
Correct?}{\PRL{79}{1997}{3577},} {\hh 9710009.}}

\lref\sen{\paper {A.~Sen}{$D0$ Branes on $T^n$ and Matrix Theory}{Adv.
Theor. Math. Phys.~{\bf 2} (1998) 51,} {\hh 9709220.}}

\lref\laroche{\paper {L.~Baulieu, C.~Laroche} {On Generalized Self-Duality
Equations Towards Supersymmetric   Quantum Field Theories Of
Forms}{\MPL{A13}{1998}{1115},}{\hh  9801014.}}

\lref\bsv{\paper {M.~Bershadsky, V.~Sadov, C.~Vafa} {$D$-Branes and
Topological Field Theories}{\NPB{463} {1996}{420},}{\hh9511222.}}

\lref\vafapuzz{\paper {C.~Vafa}{Puzzles at Large N}{}{\hh 9804172.}}

\lref\dvv{\paper {R.~Dijkgraaf, E.~Verlinde, H.~Verlinde} {Matrix String
Theory}{\NPB{500}{1997}{43},} {\hh9703030.}}

\lref\wynter{\paper {T.~Wynter}{Gauge Fields and Interactions in Matrix
String Theory}{\PLB{415}{1997}{349},}{\hh9709029.}}

\lref\kvh{\paper {I.~Kostov, P.~Vanhove}{Matrix String Partition
Functions}{}{\hh9809130.}}

\lref\ikkt{\paper {N.~Ishibashi, H.~Kawai, Y.~Kitazawa, A.~Tsuchiya} {A
Large $N$ Reduced Model as Superstring}{\NPB{498} {1997}{467},}{\hh
9612115.}}

\lref\ss{\paper {S.~Sethi, M.~Stern} {$D$-Brane Bound States
Redux}{\CMP{194}{1998} {675},}{\hh 9705046.}}

\lref\mns{\paper {G.~Moore, N.~Nekrasov, S.~Shatashvili} {$D$-particle Bound
States and Generalized Instantons}{} {\hh 9803265.}}

\lref\bsh{\paper {L.~Baulieu, S.~Shatashvili} {Duality from Topological
Symmetry}{} {\hh 9811198.}}

\lref\pawu{ {G.~Parisi, Y.S.~Wu} {}{ Sci. Sinica  {\bf 24} {(1981)} {484}.}}

\lref\coulomb{ {L.~Baulieu, D.~Zwanziger, }   {\it Renormalizable Non-Covariant
Gauges and Coulomb Gauge Limit}, {Nucl.Phys. B {\bf 548 } (1999) 527.} {\hh
9807024}.}

\lref\rcoulomb{ {D.~Zwanziger, }   {\it Renormalization in the Coulomb
gauge and order parameter for confinement in QCD}, {Nucl.Phys. B {\bf 538
} (1998) 237.} {}}

\lref\horne{ {J.H.~Horne, }   {\it
Superspace versions of Topological Theories}, {Nucl.Phys. B {\bf 318
} (1989) 22.} {}}

\lref\sto{ {S.~Ouvry, R.~Stora, P.~Van~Baal }   {\it
}, {Phys. Lett. B {\bf 220
} (1989) 159;} {}{ R.~Stora, {\it Exercises in   Equivariant Cohomology},
In Quantum Fields and Quantum Space Time, Edited
by 't Hooft et al., Plenum Press, New York, 1997}            }

\lref\dzvan{ {D.~Zwanziger, }   {\it Vanishing of zero-momentum lattice
gluon propagator and color confinement}, {Nucl.Phys. B {\bf 364 }
(1991) 127.} }

\lref\dan{ {D.~Zwanziger},  {\it Covariant Quantization of Gauge
Fields without Gribov Ambiguity}, {Nucl. Phys. B {\bf   192}, (1981)
{259}.}{}}

\lref\danzinn{  {J.~Zinn-Justin, D.~Zwanziger, } {}{Nucl. Phys. B  {\bf
295} (1988) {297}.}{}}

\lref\fiz{  {E. G. Floratos, J. Iliopoulos, D.~Zwanziger, } {A
covariant ghost-free perturbation expansioin for Yang-Mills Theories}{Nucl.
Phys. B  {\bf 241} (1984) {221}.}{}}

\lref\danlau{ {L.~Baulieu, D.~Zwanziger, } {\it Equivalence of Stochastic
Quantization and the-Popov Ansatz,
  }{Nucl. Phys. B  {\bf 193 } (1981) {163}.}{}}

\lref\munoz{ { A.~Munoz Sudupe, R. F. Alvarez-Estrada, } {}
Phys. Lett. {\bf 164} (1985) 102; {} {\bf 166B} (1986) 186. }

\lref\okano{ { K.~Okano, } {}
Nucl. Phys. {\bf B289} (1987) 109; {} Prog. Theor. Phys.
suppl. {\bf 111} (1993) 203. }

\lref\singer{
 I.M. Singer, { Comm. of Math. Phys. {\bf 60} (1978) 7.}}

\lref\neu{ {H.~Neuberger,} {Phys. Lett. B {\bf 295}
(1987) {337}.}{}}

\lref\testa{ {M.~Testa,} {}{Phys. Lett. B {\bf 429}
(1998) {349}.}{}}

\lref\Martin{ L.~Baulieu and M. Schaden, {\it Gauge Group TQFT and Improved
Perturbative Yang--Mills Theory}, {  Int. Jour. Mod.  Phys. A {\bf  13}
(1998) 985},   hep-th/9601039.}

\lref\baugros{ {L.~Baulieu, B.~Grossman, } {\it A topological Interpretation
of  Stochastic Quantization} {Phys. Lett. B {\bf  212} {(1988)} {351}.}}

\lref\bautop{ {L.~Baulieu}{ \it Stochastic and Topological Field Theories},
{Phys. Lett. B {\bf   232} (1989) {479}}{}; {}{ \it Topological Field Theories
And Gauge Invariance in Stochastic Quantization}, {Int. Jour. Mod.  Phys. A
{\bf  6} (1991) {2793}.}{}}

\lref\bautopr{  {L.~Baulieu, B.~Grossman, } {\it A topological Interpretation
of  Stochastic Quantization} {Phys. Lett. B {\bf  212} {(1988)} {351}};
 {L.~Baulieu}{ \it Stochastic and Topological Field Theories},
{Phys. Lett. B {\bf   232} (1989) {479}}{}; {}{ \it Topological Field Theories
And Gauge Invariance in Stochastic Quantization}, {Int. Jour. Mod.  Phys. A
{\bf  6} (1991) {2793}.}{}}

\lref\samson{ {L.~Baulieu, S.L.~Shatashvili, { \it Duality from Topological
Symmetry}, {JHEP {\bf 9903} (1999) 011, hep-th/9811198.}}}{}

\lref\halpern{ {H.S.~Chan, M.B.~Halpern}{}, {Phys. Rev. D {\bf   33} (1986)
{540}.}}

\lref\yue{ {Yue-Yu}, {Phys. Rev. D {\bf   33} (1989) {540}.}}

\lref\neuberger{ {H.~Neuberger,} {\it Non-perturbative gauge Invariance},
{ Phys. Lett. B {\bf 175} (1986) {69}.}{}}

\lref\gribov{  {V.N.~Gribov,} {}{Nucl. Phys. B {\bf 139} (1978) {1}.}{}}

\lref\huffel{ {P.H.~Daamgard, H. Huffel},  {}{Phys. Rep. {\bf 152} (1987)
{227}.}{}}

\lref\damhuff{ {P.H.~Daamgard, H.~Huffel, Eds.},  {Stochastic
Quantization, } World Scientific (1988).}

\lref\namok{  {M.~Namiki and K. Okano, Eds,} {}{Prog. Theor. Phys. Suppl {\bf
111} (1993). {}}{}}

\lref\creutz{ {M.~Creutz},  {\it Quarks, Gluons and  Lattices, }  Cambridge
University Press 1983, pp 101-107.}

\lref\zinn{ {J. ~Zinn-Justin, }  {Nucl. Phys. B {\bf  275} (1986) {135}.}}

\lref\shamir{  {Y.~Shamir,  } {\it Lattice Chiral Fermions
  }{ Nucl.  Phys.  Proc.  Suppl.  {\bf } 47 (1996) 212,  hep-lat/9509023;
V.~Furman, Y.~Shamir, Nucl.Phys. B {\bf 439 } (1995), hep-lat/9405004.}}

 \lref\kaplan{ {D.B.~Kaplan, }  {\it A Method for Simulating Chiral
Fermions on the Lattice,} Phys. Lett. B {\bf 288} (1992) 342; {\it Chiral
Fermions on the Lattice,}  {  Nucl. Phys. B, Proc. Suppl.  {\bf 30} (1993)
597.}}

\lref\neubergerr{ {H.~Neuberger, } {\it Chirality on the Lattice},
hep-lat/9808036.}

\lref\zbgr {L.~Baulieu and D. Zwanziger, {\it QCD$_4$ From a
Five-Dimensional Point of View},    hep-th/9909006.}

\lref\books{ O.~Piguet and A.~Rouet, {\it Symmetries In Perturbative
Quantum Field Theory,} Phys.\ Rept.\  {\bf 76} (1981) 1;
C.~Becchi, {\it `Lectures On The Renormalization Of Gauge Theories,}
{\it  In Les Houches 1983, Proceedings, Relativity, Groups and Topology,
II, 787-821}; L.~Baulieu, {\it Perturbative Gauge Theories,} Phys.\ Rept.\
{\bf 129}
(1985) 1; O.~Piguet and S.~P.~Sorella,
{\it Algebraic renormalization: Perturbative renormalization, symmetries
and anomalies, } {\it  Berlin, Germany: Springer (1995) (Lecture notes in
physics: m28)}.}

\lref\nie{N. K. Nielsen, {\it On The Gauge Dependence Of Spontaneous
Symmetry Breaking In Gauge Theories},
Nucl.\ Phys.\ {\bf B101}, 173 (1975),
H. Kluberg-Stern and J. B. Zuber, {\it Renormalization Of Nonabelian Gauge
Theories In A Background Field Gauge. 2. Gauge Invariant Operators},
Phys.\ Rev.\  {\bf D12}, 3159 (1975); {\it Renormalization Of Nonabelian
Gauge Theories In A Background Field Gauge: 1. Green Functions},
Phys.\ Rev.\  {\bf D12}, 467 (1975); {\it Ward Identities And Some Clues To
The Renormalization Of Gauge Invariant Operators}, {\bf D12}, 482 (1975);
O. Piguet and K. Sibold, {\it Gauge Independence In Ordinary Yang-Mills
Theories},
Nucl.\ Phys.\ B {\bf 253}, 517 (1985); O.~M.~Del Cima, D.~H.~Franco and O.~Piguet,
{\it Gauge independence of the effective potential revisited}
Nucl.\ Phys.\  B {\bf 551}, 813 (1999),
P. Gambino and P. A. Grassi, {\it The Nielsen identities of the SM and the
definition of mass},
hep-ph/9907254, to appear in Phys.\ Rev.\ D.}

\lref\bla{A.~Blasi, O.~Piguet and S.~P.~Sorella,

{\it Landau gauge and finiteness,} Nucl.\ Phys.\  {\bf B356} (1991) 154.}%

\lref\grassi{P.~A.~Grassi,

{\it The Abelian anti-ghost equation for the standard model in the 't Hooft
background gauge,} Nucl.\ Phys.\  {\bf B537} (1999) 527.{}}



\Title{\vbox
{\baselineskip 10pt
\hbox{hep-th/0006036}
\hbox{LPTHE-00-20}
\hbox{NYU-TH-30.5.00}
 \hbox{   }
}}
{\vbox{\vskip -30 true pt
\centerline{
   }
\medskip
 \centerline{Gauge and Topological Symmetries}
\centerline{in the Bulk Quantization of
Gauge Theories}
\medskip
\vskip4pt }}
\centerline{{\bf Laurent Baulieu}$^{\star \dag  \S    }$,
{\bf
Pietro Antonio Grassi}$^{ \ddag}$  and  {\bf  Daniel
Zwanziger}$^{ \ddag}$}
\centerline{baulieu@lpthe.jussieu.fr, Daniel.Zwanziger@nyu.edu,
pag5@nyu.edu}
\vskip 0.5cm
\centerline{\it $^{\star}$LPTHE, Universit{\'e}s P. \& M. Curie (Paris~VI) et
D. Diderot (Paris~VII), Paris,  France,}

\centerline{\it $^{\dag}$ Enrico Fermi Inst. and Dept. of Physics, University
of Chicago, Chicago, IL 60637, USA }
\centerline{\it $^{\ddag}$   Physics Department, New York University,
New-York,  NY 10003,  USA}

\def\WW{w}

\medskip
\vskip  1cm
\noindent
 A gauge theory with 4 physical dimensions can be consistently
expressed as a renormalizable
topological quantum field theory in 5 dimensions.  We extend the symmetries
in the 5-dimensional framework to include not only a topological BRST operator
$s$ that encodes the invisibility of the ``bulk" (the fifth dimension), but
also a
gauge BRST operator $\WW$ that encodes gauge-invariance and selects
observables.  These symmetries provide a rich structure of Ward identities
which assure the renormalizability of the theory, including
non-renormalization theorems.  The 5-dimensional approach considerably
simplifies conceptual questions such as for instance the Gribov
phenomenon and fermion doubling.  A confinement scenario in the
5-dimensional framework is sketched.  We detail the five-dimensional
mechanism of anomalies, and we exhibit a natural lattice discretization that is
free of fermion doubling.

\Date{\ }

\def\e{\epsilon}
\def\demi{{1\over 2}}

\def\pa{\partial}
\def\a{\alpha}
\def\b{\beta}
\def\d{\delta}
\def\c{\gamma}
\def\m{\mu}
\def\n{\nu}
\def\r{\rho}

\def\l{\lambda}
\def\L{\Lambda}

\def\P{\Psi}
\def\F{\Phi}

\def\WW{w}
\newsec{Introduction}

 In \zbgr\ we gave arguments for the relevance of a  five-dimensional
representation of gauge theories in four-physical dimensions. We proved
the existence of a local quantum field theory that is  perturbatively
renormalizable  by power counting  and  free of the Gribov ambiguity.  We
also gave its  lattice formulation, and suggested that the 5-dimensional
framework naturally avoids fermion doubling.  A possible interpretation is
that the fifth-dimension is  the stochastic time that  Parisi and Wu proposed a
long time ago for stochastically quantizing  the Yang--Mills theory
\pawu, and the 4-dimensional physical theory lives in any given chosen
time-slice of the  space with five dimensions.  However the beauty of the
resulting theory suggests that the fifth time plays a more fundamental and
more general role for example in elucidating the Gribov and fermion doubling
problems.  In particular we consider the 5-dimensional functional integral to
be more fundamental than the Langevin or Fokker-Planck equations.

After the discovery of the four-dimensional Yang--Mills
 topological quantum field theories
\wittentopo,  it was  realized that the supersymmetric formulation
of  stochastic quantization also determines  a topological field theory
\bautopr. This was an early example of a  physical theory that lives in the
``boundary'' of a space with an additional dimension,
independently of the process in the ``bulk'' that
determines the many possible ways the a Fokker--Planck distribution
converges to an equilibrium distribution.  One recognizes a
holographic phenomenon at work here, and
various connections between different theories have been exhibited in
this way, with the idea that stochastic quantization is analogous to a
Stokes theorem for the path integral \baex.  With the inclusion of fermions,
and BRST ghosts, the local 5-dimensional formulation transcends its
purely stochastic origin.   From now on, we call it bulk quantization.

        One must establish the consistency of quantization with an
additional time,
and the four-dimensional Yang--Mills theory is an important and
challenging case. Long before the invention of topological field theory, it
was shown that the Parisi--Wu conjecture
\pawu\ is compatible at the perturbative level with the Faddeev--Popov
method \danlau.  For
reviews of stochastic quantization, see  \huffel, \damhuff\ and \namok.
Further development of stochastic quantization, particularly gauge-invariant
stochastic regularization, may be found in \bautopr, \halpern\ and \yue, and
for renormalization in \danzinn.  For renormalization of non-gauge theories
in the 5-dimensional formulation, see \zinn.

The hope of enriching our perspectives is of course at the
 non-perturbative level.  The 5-dimensional theory possesses a
 supersymmetry of the topological type which ensures that the
 expectation-value, once equilibrium is achieved, is independent of the
 details of the initial conditions.  The existence of an
 unobservable fifth time considerably simplifies many conceptual
 problems that perplex the sole and too narrow four-dimensional
 perspective.  For example the local quantum field theory in 5
 dimensions avoids the question of Gribov copies that jeopardizes the
 Faddeev--Popov prescription in four dimensions \dan.  As shall be discussed
elsewhere, the introduction of the additional time allows one to replace
problematic gauge fixing by an appropriate {\it gauge transformation} in such a
way that the Gribov question becomes irrelevant. The idea of looking at gauge
theories from five dimensions can be put in correspondence with the
description of conformal theories from the Chern--Simons action in three
dimensions.

  Our definition of observables in \zbgr\ was not entirely
 satisfactory to the extent that it was not based on an invariance
 principle.  Here we will fill this hole in our presentation, and
 introduce, in  addition to the supersymmetry operator $s$ that
 expresses  quantization with a fifth time \zinn, another BRST-symmetry
operator,
 $\WW $, that implements gauge symmetry in 5 dimensions and which is
 compatible with~$s$.  This necessitates introducing an additional
 field, in a way that is inspired by an idea originally due to Horne
 \horne\ in the context of the topological Yang--Mills theory in four
 dimensions, and emphasized in subsequent works using  the
 idea of equivariant cohomology \sto.  The 5-dimensional action
 of the Yang--Mills theory is entirely defined by the requirement of $\WW$
 and $s$ symmetry.  We will define observables by the cohomology of
 $\WW $ taken at a fixed but arbitrary time-slice.

	There are two features of the 5-dimensional formulation of
4-dimensional
quantum field theory -- whether of $\f^4$ or of Yang-Mills type -- which we
wish
to emphasize.  The first is that even though the action is $s$-exact,
$I = sX$, where $s$
has all possible characteristics of a topological BRST operator, we are
{\it not} in
the context of a topological theory of the usual type, as was noted in \zbgr.
The reason is that the observables $O$ are {\it not} required to be
$s$-invariant, $sO \neq 0$.  Indeed the cohomology of $s$ is empty,
so if the observables were
$s$-invariant, they would be $s$-closed, $O = sY$, and would have vanishing
expectation-value, because our theory has no zero modes
that would allow $<sY>\neq 0$.
Rather, in the case of a
scalar theory, the observables are all possible correlators taken at equal
time,
and in the case of gauge theories,  they are also required to be in the
cohomology of another BRST operator $w$, such that
$w^2=sw+ws=0$. Beyond technical details and subtleties, $w$ is the expression
of the gauge symmetry in the five-dimensional framework. The
five-dimensional theory is thus  not a topological theory,  although its
$s$-exact action looks topological. Rather, our interpretation is that the
latter
property is the simplest way to assure that possible renormalization
constants are the same as in the 4-dimensional theory.  The second
feature, which holds both for gauge and scalar theories,  is that the
correlators
are distributions in 5 dimensions, and in general they have singularities at
equal times.  As a result, physical observables, which are restricted to a time
slice, may not be well-defined.   Indeed, as shown in Appendix E, the
correlator
of three chiral currents,
$\langle j(x_1, t_1) j(x_2, t_2)j(x_3, t_3)\rangle$,
is ambiguous in the equal-time limit, and this is the origin of the triangle
anomaly in the 5-dimensional formulation. We expect that this kind of
obstruction to consistent equal-time limits for the observables occurs only
when the gauge theory is anomalous.

 We will explain in a separate paper that the condition of
fixed time relies  on the correspondence
 of Schwinger--Dyson equations  in 4 and 5 dimensions.

        The 5-dimensional  formulation accommodates in a local
description the gauges that are actually used at present in lattice gauge
theory
for numerical evaluation.  These gauges, such as the minimal Landau, minimal
Coulomb or maximal Abelian gauges, are fixed by minimizing an appropriately
chosen functional.  They cannot be correctly described by the Faddeev--Popov
method that is characterized by a local gauge condition, such as $\p_\l A_\l =
0$, that does not distinguish between minima and saddle points of the
minimizing functional.  Nevertheless these gauges are represented by a local
action in 5 dimensions.  We would like to emphasize that the 5-dimensional
local theory does not reproduce the gauge-non-invariant part of  the standard
4-dimensional Faddeev--Popov formulation.  The latter can only be reached
from the 5-dimensional theory by means of a non-local action~\danlau, and
then only at the perturbative level.

        Instead of a local gauge-fixing in 4 dimensions which is known not
to exist
\singer, the 4-dimensional probability distribution is obtained from the
solution
of a Fokker--Planck equation in 5 dimensions that preserves the
4-dimensional probability at each instant.  The solution of
the Fokker--Planck equation, which determines all gauge-invariant correlation
functions in 4 dimensions is represented by a local 5-dimensional gauge theory
of topological type.  Its  path integral formula is valid
non-perturbatively, and
we have previously given its BRST-invariant lattice regularization~\zbgr.

        The organization of the paper is as follows.  The basic formulation
of the
theory is presented in sec.~2.  Here the field content is explained, the $s$-
and $\WW $-symmetries are defined, and the most general
renormalizable $s-$ and $\WW $-invariant action is exhibited. It is important
in this regard to keep in mind that the fifth time has engineering dimenison
double of that of ordinary space-time coordinates
$[\p/\p t] = 2 [\p/\p x_\m] = 2$.  In sec.~3,
we briefly sketch a physical interpretation of the theory.  A number of further
developments of the theory and various issues are explored in the Appendices.
In Appendix A, we derive the Ward identities which assure the stability of the
theory under renormalization and fix a number of renormalization constants.
In particular it is shown that $gA_5$ is invariant under renormalization in the
minimal Landau gauge, and thus may be used to define an invariant charge in
QCD.  In Appendix B, properties of the minimal Landau-gauge are derived.  This
gauge provides the frame for a confinement scenario that is described in
Appendix C, in which long-range ``forces" are transmitted by $A_5$.  In
Appendix D, the theory is extended to Dirac spinor fields.  The important topic
of anomalies is addressed in Appendix E.  Here a 5-form is found which is a
candidate for an obstruction in the 5-dimensional theory.  We show however
that the familiar triangle anomaly of the 4-dimensional theory has another
origin: it is a singularity that appears when fifth times are set equal, as is
necessary to obtain the physical 4-dimensional correlators.  In Appendix F, we
show that fermion doubling may be avoided by lattice discretization of the
5-dimensional theory.

 \newsec{Field content and symmetries}
\def\tr{{\rm tr}}
\def\lb{{\bar{\lambda}}}
\def\mb{{\bar{\mu}}}
\def\eb{{\bar{\eta}}}
\def\Fb{{\bar{\Phi}}}
\def\Pb{{\bar{\Psi}}}
\def\cb{{\bar{c}}}

We first focus on the pure Yang--Mills case, and will introduce coupling to
spinors later.  The 5-dimensional action used in \zbgr\
\eqn\tan{\eqalign{ I =
 \int dx^\m dx^5\ s\ \Tr \big(
\bar \P^\m   ( F_{5\m} - D_\l F_{\l\m} -\demi b_\m) + ...
 \big) }}
is $s$-exact, and $...$ will be specified shortly.
\tan\ looks like a topological action. Indeed,
  $s$ is a topological
BRST operator that will be defined shortly, as well as all relevant fields.
Greek indices denote 4-dimensional Euclidean components,
$\l, \m = 1,...4$.   The first term in
\tan\ is invariant under 5-dimensional gauge transformations.  Roughly
speaking, it  concentrates the path integral around the solutions of the
equation
$F_{5\m} - D_\l F_{\l\m} =0$.  The
five-dimensional gauge symmetry of the action will be broken in a
BRST-invariant way by means of
\eqn\gfixing{\eqalign{ aA_ 5 = \pa_\l A_\l \ ,}}
where $a$ is a gauge parameter.
This condition should not be interpreted as a Faddeev--Popov gauge-fixing.
As will be discussed elsewhere, $A_5 = \dot{g}g^{-1}$ is the generator of a
time-dependent gauge transformation
$g(x, t)$ that acts on $A_\m$ for $\m = 1, ...4$, , and as such it can be fixed
arbitrarily, apart from the constraints imposed by renormalizability.  Indeed
the role of
$A_5$ as the generator of a time-dependent gauge transformation survives
the algebraic quantization of the 5-dimensional theory, as is shown in
Appendix A.  The successful avoidance of the Gribov problem is reflected in
the fact that the ghost propagators are parabolic  when \gfixing\  is enforced
in a BRST-invariant way, so the ghost propagators are retarded, $G(x, t) = 0$
for $t < 0$, and  consequently closed ghost loops vanish and the
Faddeev--Popov determinant is trivial.  Heuristically, one recognizes that
\gfixing\  is like an axial gauge in $A_5$, for which there is no Gribov
ambiguity, and the infinite range of the variable $t = x_5$ avoids the Singer
theorem \singer.

 When the ghost and auxiliary fields are
integrated out, one obtains the action of stochastically quantized gauge
theory \pawu,
\eqn\tgfa{\eqalign{ I= &
 \int dx^\m dx^5\  \Tr \big(
 F_{5\m}^2 +(D_\l F_{\l\m})^2 \big)}}
with stochastic gauge-fixing \dan.

The observables of the theory were defined in \zbgr\ in the following
intuitive way: they are correlation functions of gauge-invariant
functions of the gauge field components, $A_\m$, for $\m = 1,...4$,
taken at equal values of $t = x_5$. The Green functions can be first
computed and renormalized at different values of time, and then one
takes the limit of equal times.  From the point of view of quantum
field theory the fifth time appears as a regulator  and from a
geometrical point of view of topology, it appears as
a variable that enlarges the space and
simplifies topological properties.
 We shall refine the  definition of observables  here, and define
observables from a symmetry principle, in order  to  have a better
control of their renormalization properties. This will lead us to refine
our knowledge of the symmetry of \tan.  We want to make this notion
precise, and end up with the definition of all observables in the cohomology
of a certain symmetry operator, $\WW $, which is compatible with $s$.

 We have shown in \zbgr\ that the action \tan\ gives a perturbatively
renormalizable theory.  Although the action \tan\ is not $SO(5)$
invariant, its BRST symmetry is $SO(5)$ invariant, which is sufficient
for a consistent description, since the five-dimensional description
is holographic, and the only things that matters is to recovers the
$SO(4)$ invariance for the observables.  Power-counting constrains the
way the invariance under $SO(5)$ symmetry is reduced down to $SO(4)$:
indeed, the canonical dimension in mass units of each one of the
four-dimensional components of the gauge field is unity whereas $A_5$
has dimension two.  (The Yang--Mills coupling constant has dimension
zero).

Let us  now define all the fields, which depend  on $x^\mu$ and $t = x_5$
and are Lie algebra valued. New fields will occur, as
compared to  \zbgr\ because of the  $\WW $-symmetry.
 The bose and fermi fields are in one-to-one
correspondence, and are related by the topological BRST  operator $s$, as in
a supersymmetric theory.  (This suggests the possibility of a  link between
the topological BRST operator $s$ and some kind of Poincar\'e twisted
supersymmetry.)  In addition to $A_\m$ and $A_5$ there are:
$c$, which has the quantum numbers the ordinary Faddeev--Popov ghost
but plays a different role; $\Psi_\m$ and $\Psi_5$, the topological Fermi
vector ghosts corresponding to $A_\m$ and $A_5$; and
$\Phi$, a commuting  ghost of ghost. These fields have respectively  ghost
number $N_s = 1,1,1$ and $2$.    We also have corresponding anti-ghosts
$\cb$, $\Pb_\m$ and $\Fb$, with ghost number  $N_s = -1,-1$ and $-2$
respectively, and Lagrange multipliers
$b_\m$, $l$ and $\eb$, with ghost number
$N_s = 0,0$ and $-1$ respectively.  (In \zbgr\  we used $\cb = \Pb_5$ and
$l = b_5$.)  See the diagram below for the relations of these fields and their
quantum numbers.

        There are five degrees of freedom for the choice of a dynamics, one for
each component of $A$.  The Lagrange multiplier field $l$ serves to
impose the condition \gfixing, and the
$b_\mu$'s  enforce the Langevin equation.  Finally $ \eb$ is a fermion
Lagrange multiplier for the gauge fixing of the longitudinal modes in
$\Psi$.

The topological BRST operator $s$ is not relevant for defining
observables.  Indeed, the cohomology of $s$ with ghost number zero is
empty.  Although it involves  the gauge symmetry with a ghost of ghost
phenomenon to ensure its nilpotency, it only
  represents  the irrelevance of the
  details of the process in the bulk.

  To encode gauge-invariance and distinguish gauge-invariant
observables, we need another BRST operator, which we call $\WW $.  We  thus
introduce a  second  ghost number
$N_w$ and new fields.   The total grading is the sum $N_s + N_w + p$,
where $p$ is the ordinary form degree.  We need a second Faddeev--Popov
ghost $\l$ and, as will be explained shortly, also its ghost of ghost $\m$.
The
ghost-field $\l$ plays the role of the ordinary Faddeev--Popov ghost, and in
the ``effective'' theory on the boundary, the $\WW $ operator induces the
ordinary BRST symmetry.   One actually has a quartet of additional ghosts
and anti-ghosts
$\l$, $ \m$, $\lb$, $\mb$, which maintains supersymmetry.   All the
fields  and ghosts  besides $\l$, $ \m$, $\lb$ and $\mb$
have ghost number
$N_w=0$.  The latter have respectively ghost number
$N_w=1,1,-1,-1$, and $N_s=0,1,0,-1$.  It is convenient to use a
bigrading  notation $\phi^{(N_s,N_w)}$ that indicates the ghost numbers
of any given field $\phi$.  With this notation one has
 $(s\phi)^{(N_s+1,N_w)}$ and  $(\WW \phi)^{(N_s,N_w+1)}$,
$(\WW s\phi)^{(N_s+1,N_w+1)}$.  All ghost numbers can be read from the
following diagram:

\eqn\diagram{\eqalign{\matrix{
                        \matrix{&&&&A^{(00)}_\m,A^{(00)}_5\cr
                       &&&\swarrow&&\ \cr
                    &&\Psi_\mu^{(1,0)},\Psi_5^{(1,0)}
&&&&\bar{\Psi}_\m^{(-1,0)}\cr
                    &&  c^{(1,0)},\l^{(0,1)}
&&&& \bar{c} ^{(-1,0)}
\cr
                    && \
&&&& \bar{\l} ^{(0,-1)}
 \cr
                       &\swarrow&&\ &&\swarrow&&\nwarrow \cr
                       \Phi^{(2,0)},\mu^{(1,1)}
&&&&
b_\mu^{(0,0)}&&&&\bar{\Phi}^{(-2,0)},\bar{\mu}^{(-1,-1)}\cr
%
%
%
&&&&
l ^{(0,0)}&&&&\
\cr
&\ &&\ &&\ &&\swarrow\cr
\cr
                    &&\
&&&&
\bar \eta ^{(-1,0)}\cr
                         && \ &&&& \cr} }}}
The engineering dimensions (in mass units) of the fields are assigned according
to:
\eqn\candim{\eqalign{
& [c] = [\l] =  [\m] =  [\F] = 0   \cr
& [{{\partial }\over{\partial x^\mu}}]=[A_\m] =  [\P_\m] = 1  \cr
& [{{\partial }\over{\partial x^5}}]=[A_5] =  [\P_5] = 2  \cr
& [b_\m] = [\Pb_\m] = 3   \cr
& [l] = [\cb] =  [\mb] =  [\lb] = [\Fb] = [ \eb] = 4. }}
These values will play a key role in the determination
of the five-dimensional action. The asymetry  in the dimensions of
the fields with ordinary labels $\mu$ and $5$ is absolutely crucial, and will
explain how the good properties of the power counting in the conventional
four-dimensional formulation of gauge theories are still present in the
five-dimensional formulation.  Moreover, since the anti-ghosts have
dimension 3 or 4 and $\p / \p t$ has dimension 2, all ghost actions will
be parabolic and all ghost propagators will be retarded.

In
\diagram, the arrows conveniently relate the fields that can be transformed
into
each other by
$s$- or  $\WW
$-transformations that we will give shortly, in accordance with the separate
conservation of the ghost numbers $N_s$ and $N_w$. The diagram \diagram\
exhibits the boson-fermi pairing which always occurs in theories with a
``topological'' symmetry. The subtlety is the way the symmetry is
expressed,  with a delicate separation of the gauge symmetry
transformations from the topological transformations, expressed respectively
by the $\WW $- and $s$-transformation.  By convention, since the 2 ghost
numbers are conserved, we assign dimension in such a way that both $s$ and
$\WW $ leave dimension unchanged.

        We now define the action of $s$ and $\WW $.\foot{A more symmetric
diagram would involve the idea of anti-$s$ and anti-$\WW $ symmetries}  We
require
\eqn\stable{\eqalign{ \WW ^2=s^2=s\WW +\WW s=0 .}}
This property is  automatically assured by   defining the symmetry in the
following geometrical way, which  is by now a
standard in the BRST paradigm:
\eqn\gfca{\eqalign{ &( s +\WW +d)(A+c+\l)+\demi[A+c+\l,A+c+\l]=
F+  \P_\m dx^\m+  \P_ 5  dx^ 5+  \F    \cr
& (s+\WW +d)(F+  \P_\m dx^\m+  \P_ 5  dx^ 5+  \F  )  =- [A+c+\l,
 F+  \P_\m dx^\m+ \P_ 5  dx^ 5+  \F   ]  \cr
    }}
The separate conservation of the ghost numbers $N_s$ and $N_w$
determines the action of $s$ and $\WW $ on all fields except
$s\l$ and $  \WW  c$, since the  geometrical  equation only determines
\eqn\undet{\eqalign{
s\l +\WW  c=-[\l,c].
}}
  To resolve this degeneracy we introduce the ghost
$\mu$, with $N_s= N_w=1$, and impose
$s\l=\mu$.  The result of the decomposition is:
\def\ss{s}
 \eqn\defta {\eqalign{   \ss   A _{ \mu } &=    \Psi_\m +  D _{ \mu }c ,
~~~\ss    \Psi_{\mu } = D _\m \Phi
  -[c ,\Psi_\mu], ~~~\ss    \, c  =    \Phi -{1\over 2}[ c ,c ] \ ,
\cr
 \ss   A _{ 5 } &=    \Psi_5+  D_{ 5 }c , ~~~\ss    \Psi _{5 } = D _5
\Phi -[c ,\Psi_5], ~~~
\ss   \Phi =  - [c  ,\Phi] \  .
 ~~~\cr
 &~~~~~~~~~~~~~~~~~~~~~~~\ss  \l   =   \mu    , ~~~\ss   \mu = 0 \  }}

\def\ss{\WW }
\eqn\defta {\eqalign{     \WW   A _{ \mu } &=   D _{ \mu }\l , ~~~   \WW
\Psi_{\mu } =
  -[\l ,\Psi_\mu], ~~~ \WW       c  = -\mu  - [ \l ,c ] \ ,
\cr
 \WW   A _{ 5 } &=   D _{ 5 }\l , ~~~
 \WW    \Psi _{ 5 }  =-[\l ,\Psi_5], ~~~
 \WW    \Phi =  - [\l  ,\Phi]   ~~~
\cr&~~~~~~~~~~ \WW \l   = -\demi [ \l ,\l ]   ~~~ \WW    \mu   =  - [\l
,\mu] }}
Observe that $\WW $ acts on $A_\m$ and $A_5$ like an infinitesimal
gauge transformation with gauge parameter $\l$, and observables will be
required to be $\WW $-invariant.  Because of the inhomogeneous term
$\m$ in $\WW c=-\mu - [\l, c]$,  $\WW $-invariance of a quantity with
ghost-number zero assures  that it is independent of $c$.  Moreover $\mu$  is
a topological ghost for $c$ and consequently $\WW $-invariant observables are
independent of $c$.

        For the anti-ghosts of the $s$-symmetry, $\bar \Psi_\m$, $\bar \Phi$,
and the corresponding Lagrange multipliers
$ b_\m$ and $\eb$, one has:
\eqn\gfca{\eqalign{ ( s +\WW  )\bar \Psi_\m+ [ c+\l,\bar \Psi_\m]&= b_\m \cr
 (s+\WW  )b_\m + [ c+\l, b _\m  ] & =   [ \Phi ,
\bar \Psi_\m    ]\cr
    }}
\eqn\gfca{\eqalign{ ( s +\WW  )\bar \Phi + [ c+\l,\bar\Phi ]&=
\eb \cr
 (s+\WW  )\eb + [ c+\l,
\eb  ] & =   [ \Phi ,
\bar\Phi     ]\cr
    }}
(Notice  that
$(s+\WW )^2=0$ amounts to $(s+\WW +[c = \l,{\bf .}])^2=[\Phi,{\bf .}]$)  This
definition implies:
\eqn\gfca{\eqalign{   s \bar \Psi_\m=- [ c,\bar \Psi_\m]+ b_\m , ~~~  s
b_\m=-   [ c , b _\m  ] +   [ \Phi,
\bar \Psi_\m    ]
    }}
\eqn\gfca{\eqalign{   \WW   \bar \Psi_\m=- [  \l,\bar \Psi_\m] , ~~~
 \WW   b_\m =- [  \l, b _\m  ]   ]
    }}
\eqn\gfca{\eqalign{   s \bar \Phi=- [ c,\bar \Phi]+\eb , ~~~
  s\eb =-   [ c ,
\eb ] +   [ \Phi,
\bar \Phi    ]
    }}
\eqn\gfca{\eqalign{   \WW  \bar \Phi=- [  \l,\bar \Phi] , ~~~
 \WW  \eb  =- [  \l,
\eb  ]    \cr
    }}

Finally we give the action of $\WW $ and $s$ on the anti-ghosts and
Lagrange multipliers $\bar c$,  $l $, $\bar \mu$ and $\bar\l$. We
take:
\eqn\gfca{\eqalign{ ( s +\WW  )\bar \mu   =
\bar c+\bar\l \cr
 ( s +\WW  )(\bar c  + \bar\l )  =0
    }}
In order to break the degeneracy of the geometrical equation
$\WW \bar\l+s \bar c=0$, we introduce the new Lagrange multiplier field $l$ ,
and impose $s \bar c = -l$.  After expansion in the ghost numbers $N_s$ and
$N_w$ we obtain,
\eqn\gfca{\eqalign{   s \bar c=-
l  , ~~~  s l =   0\cr
 s \bar \mu=
\lb , ~~~  s \lb=   0
 }}
\eqn\gfca{\eqalign{   \WW \bar\mu& =\bar c, ~~~  \WW  \bar c=   0\cr
\WW  \bar \l&=
l  , ~~~ \WW  l =   0
}}

We are now ready to ask the standard question, within the BRST
paradigm, of determining the gauge-fixed action, the possible
counter-terms, and the anomalies of a theory that is gauge and Lorentz
invariant and renormalizable by power counting.  Here we have the two
symmetries $\WW $ and $s$, with
$(\WW +s)^2=0$, and we must classify the  local functions of the
fields that are both $\WW $- and $s$-invariant. The important question of
anomalies is addressed in Appendix E, where we display an intriguing new
cocycle stemming from \gfca.   In what follows we consider the question of
determining the action in 5 dimensions from the requirement of
$s$- and $\WW $-symmetry.

	We wish to find the most general solution to the equations $sI = 0$
and $wI
= 0$, where $I = \int dt d^4x \cal{L}$, and  $\cal{L}$ is a local Lagrangian
density of engineering dimension 6, and $I= I^{(0, 0)}$.  Here we use the
notation defined above to indicate the $N_s$ and $N_w$ ghost quantum
numbers.  We rely on the fact that the cohomology of $s$ with ghost number
zero is empty.  Thus the action $I^{(0,0)}$ must be an $s$-exact term,
$I^{(0, 0)} = sI^{(-1, 0)}$.  We shall use the strategy of descent equations,
with $d$ and $s$  replaced respectively by $s$ and  $w$, because our~$s$,
like~$d$, has empty cohomology.  By $w$-invariance of $I^{(0,0)}$ we have
$0 = wI^{(0, 0)} = wsI^{(-1, 0)} = - swI^{(-1, 0)}$.  Since the cohomology of
$s$ is empty, this gives
\eqn\descenta{\eqalign{
wI^{(-1, 0)} = sI^{(-2, 1)}.
}}
Upon multiplying this equation by $w$, we obtain
$0 = wsI^{(-2, 1)} = - swI^{(-2, 1)}$, and so, since the cohomology of $s$ is
empty, we have
\eqn\descentb{\eqalign{
wI^{(-2, 1)} = sI^{(-3, 2)}.
}}

	Now we use the fact that we have assigned the engineering
dimensions to the
fields in such a way that the operators $s$ and $w$ preserve engineering
dimension, so the $I^{(i,j)}$ that we have introduced all have engineering
dimension $[I^{(i,j)}] = 0$, and all the corresponding densities have
engineering dimension 6.  Since $I^{(-3, 2)}$ has ghost number $N_s = -3$, it
must contain either at least 3 factors of the anti-ghost fields with
$N_s = -1$, or else at least one power of the anti-ghost field
$\bar{\F}^{(-2,0)}$ and another anti-ghost field.  In the first case the
corresponding density has engineering dimensions 9 or greater because
all anti-ghost fields have dimension 3 or 4, as one sees from \candim.
In the second case the corresponding density has
engineering dimension 7 or greater.  We conclude that
$I^{(-3, 2)} = 0$, so from \descentb, we have
$wI^{(-2, 1)} = 0$.  Thus $I^{(-2, 1)}$ is of the form
\eqn\cohow{\eqalign{
I^{(-2, 1)} = I_{\rm inv}^{(-2, 1)} + w I^{(-2, 0)},
}}
where $I_{\rm inv}^{(-2, 1)}$ is an element of the cohomology of $w$.
However there is no local density in the cohomology of $w$, of engineering
dimension 6, with these ghost quantum numbers.  For $I_{\rm inv}^{(-2, 1)}$
must contain at least one power of either
$\m^{(1,1)}$ or $\l^{(0,1)}$.  If it contains $\m^{(1,1)}$, then
it must contain so many anti-ghost fields that its engineering
dimension is too high.  On the other hand if it contains $\l^{(0,1)}$,
then there is no
such element of the cohomology of $w$.  (For example
$(D_5\bar{\F})^a \l^a$ has the right dimension, but is not in the
cohomology of $w$.)  We conclude that $I_{\rm inv}^{(-2, 1)} = 0$,
so from \cohow\ we obtain
$I^{(-2, 1)} = w I^{(-2, 0)}$.  We substitute this into \descenta\ and obtain
$wI^{(-1, 0)} = swI^{(-2, 0)}$, or
\eqn\wxez{\eqalign{
w(I^{(-1, 0)} + sI^{(-2, 0)}) = 0.
}}
Thus the quantity in parenthesis is of the form
\eqn\cohowa{\eqalign{
I^{(-1, 0)} + sI^{(-2, 0)} = I_{\rm inv}^{(-1, 0)} + wI^{(-1, -1)},
}}
where $I_{\rm inv}^{(-1, 0)}$ is an element of the cohomology of $w$.  Upon
substitution of this equation into $I^{(0,0)} = sI^{(-1,0)}$, we obtain
\eqn\cohoso{\eqalign{
I^{(0, 0)}= sI_{\rm inv}^{(-1, 0)} + swI^{(-1, -1)}.
}}

It follows that the  Lagrangian density must be of the form
\eqn\dec{\eqalign{
  \ s\ \Tr [\bar \P^\m   K_\mu^{(0,0)}(A,b_\m)
+\bar \Phi    K^{(1,0)}(\Psi,A)]
+ ws\ \Tr [\bar \m  L^{(0,0)}(A )],
 }}
where $K_\mu^{(0,0)}(A,b_\m)$,
$K^{(1,0)}(\Psi,A)$ and $L^{(0,0)}(A )$ have dimension 3, 2 and 2
respectively.  Here we have used the fact that the fields
$c$, $\bar{\l}$ and and $\bar{\m}$ undergo transformations of
topological type under the infinitesimal gauge transformation $w$, so they
cannot  appear in the cohomological term $I_{\rm inv}^{(-1, 0)}$.
In fact $I_{\rm inv}^{(-1, 0)}$ must be constructed out of combinations of
local fields that transform covariantly under $w$.
Note that the only possible \WW -exact term that is
$s$-invariant and thus $s$-exact  is $s\WW $-exact.  To obtain this result  we
used ghost number conservation and power counting arguments for a local
Lagrangian density of engineering dimension 6.

	One gets by inspection that the two  first term in
\dec,   up to multiplicative  renormalization constants,
must be:
\eqn\tinv{\eqalign{  I_{\rm inv  }=
 \int dx^\m dx^5\ s\ \Tr [ \
\bar \P^\m   ( F_{5\m} - D_\l F_{\l\m} -\demi b_\m)
+\bar \F    (a' \P_5 - D_\l \P _\l ) \ ]. }}
This determines the gauge-invariant part of the action, or more precisely the
part of the action that is in the non-trival part of the cohomology of $w$ with
ghost number zero and dimension 6.  The first term, corresponds to the
5-dimensional action of stochastic quantization that one can guess from the
Langevin equation for the  four-dimensional classical Yang--Mills action. It
may be written as:
\eqn\tinva{\eqalign{  I_1=
 \int dx^\m dx^5\ s\ \Tr [
\bar \P^\m   ( F_{5\m} + {{\d S} \over {\d A_\m}} -\demi b_\m)  ],
 }}
where $S = \int d^4x (1/4) F_{\m\n}^2$ is the standard Yang--Mills action.
The second term of \tinv, that is linear in $\Fb$, fixes in a gauge-covariant
way the internal gauge invariance of
$\P_\m$ and $\Psi_5$.
These terms, that we derived in a straightforward way from the requirement of
locality  and compatibility with power counting, show the conceptual
limitation of the idea of stochastic quantization in its original formulation.
We refer to \zbgr\ for more details concerning the meaning of each term that
occurs in the expansion of \tinv.

	For a local action of dimension 6, the only possibility for
the last term in \dec\ is
\eqn\tgf{\eqalign{ I_{\rm gf}= \int dx^\m dx^5\ \WW s\ \Tr \big([\bar \mu
(aA_5 -  \partial\cdot A)] \ ,}}
and the total action is given by
\eqn\tact{\eqalign{I = I_{\rm inv  }+I_{\rm gf }  \ .  }}
 Remarkably, the only possible gauge-fixing term with dimension 6 provides a
linear gauge-fixing.  (If the coupling to matter were included, the
generalization of the Feynman--'tHooft gauge for spontaneously broken
symmetries is easily obtained, $aA_5 = \p_\m A_\m + v\f$.)  The derivation
of \tgf\ using $w$-invariance is a significant improvement compared to the
derivation in \zbgr.  The term
$I_{\rm inv} = sI_{\rm inv}^{(-1,0)}$
is the same as in \zbgr, and its expansion may be found there.

To see in detail how \tgf\ solves the question of raising the
 degeneracy with respect to ordinary   gauge transformations of \tinv, we
expand
\eqn\gfcaa{\eqalign{   \WW s[\bar \mu  (aA_5 -  \p_\n A_\n)]
= & \WW  \{ \bar \l (aA_5 - \p_\n A_\n)    \cr
& + \bar \m [ a\P_5 + aD_5c -  \p_\n (\P_\n + D_\n c)] \ \}.}}
The first term gives
\eqn\ccaa{\eqalign{  l   (aA_5 - \p_\n A_\n) - \bar \l (aD_5 \l
- \p_\n D_\n\l),}}
which fixes the gauge for $A_5$ and $\l$.  The second term
in   \gfcaa\ gives:
\eqn\caa{\eqalign{  &  \quad\quad \quad \bar c  [ a\P_5 + aD_5c - \p_\n
(\P_\n + D_\n c)]    \cr
& + \bar \m \{ a D_5 \m - \p_\n D_\n \m)
+ a[ \l ,\P_5 + D_5c] - \p_\n[\l, (\P_\n + D_\n c)] \ \} .
  }}

The equation of motion of  $\bar  c $ determines a certain  linear  combination
of
$c$ and $\P_5$.  But a similar situation occurs for   $\eb$. Using its
equation of motion  from
  $I_{\rm inv}$, $\P_5$ is determined, so both $c$ and $\P_5$ are
determined.  This can be seen by doing the translation
$\eb \to \eb -\cb$.  Finally, the  equation of motion of  $\bar \m$  determines
 $\m$, because its equation of motion is parabolic.

  An essential feature is that the equations of motion of all ghosts is
parabolic,
of the type
$(\p_t - a^{-1}\p_\n^2 + ...)\xi = 0$.  It follows that the free ghost
propagators
$G_0(x,t) = \theta(t) ({{a}\over{4 \pi t}})^2 \exp ( {{- ax^2}\over{4t}}) $
are all retarded, and consequently  {\it all closed  ghost loops vanish.}  The
only possible exception is the tadpole which gives a purely local
contribution to
the action which is a renormalization counter-term.  However even the tadpole
contribution vanishes with dimensional regularization.  With dimensional
regularization in mind, we conclude that the determinant of each ghost is
unity.

Observables $O$ are required to be in the cohomology of $\WW $, thus
$\WW O = 0$ and $O\neq \WW (X) $, with ghost number 0, and lying in a fixed
but arbitrary time slice.  However, as noted in the Introduction, they are not
$s$-invariant, $sO \neq 0$.
 We may impose additional conditions on the class of observables.
The most conservative policy would be to allow only functions of the variables
that are present in the 4-dimensional theory, namely the $A_\m$ for
$\m = 1,...4$, and Dirac spinor fields.  Expectation-values of observables are
independent of the parameters of the $\WW $-exact term, that is on the
parameter
$a$.  Their independence of the other parameter $a'$ is quite clear,
as shown in \zbgr: closed ghost loops vanish, and so cannot contribute to
the expectation values of   observables, and the later cannot depend on $a'$.
Since observables are $w$-invariant, which implies that they are gauge
invariant, we do get the observables of the 4-dimensional theory.

As for a detailed perturbative proof of the stability of
renormalization, it can be done quite rigorously, using the Ward identities
derived in Appendix A.  If one uses dimensional regularization by computing
correlators in $5-\epsilon$ dimensions, one finds that the action $I$ undergoes
a multiplication renormalization for all its fields and parameters, as a
result of
imposing  the Ward identities of both $\WW $- and
$s$ -invariance in the gauge determined by \ccaa, that is with the gauge
condition \gfixing.  It should be understood
that the correspondence between equal-time correlators in 5 dimensions and
those in 4 dimensions is for the physical observables only (defined just
above, from the $\WW $-cohomology).\foot{The comparison  for gauge
non-invariant gauge functions would require a non-local theory in five
dimensions,  \danlau, and presents no interest since locality is a key tool for
mastering perturbative renormalizability.}

\newsec  {Interpretation of the action}

        One way to exhibit the physical content of the theory is to  integrate
out all ghost
and auxiliary fields.  We will do this in two steps, in order to get an
intermediate result which will be useful later.  The first step is to
integrate out all ghost fields except $\l$ and $\lb$.   As explained above,
closed ghost loops vanish because ghost propagators are retarded,
$D(x,t) = 0$ for $t < 0$, so the ghost determinant which results from this
integration is a constant, and the action \tact\ reduces to
\eqn\action{\eqalign{I = \int d^5x [&  ib_\m (F_{5\m} - D_\l F_{\l\m})
+ (1/2)b_\m^2  \cr
& + il(aA_5 - \p_\m A_\m) + \lb(aD_5 - \p_\m D_\m)\l]
\ .}}
This action possesses the ordinary BRST invariance, here implemented by
$\WW $, that encodes gauge invariance in five dimensions.  The
$s$-invariance that stabilizes the action \tact\ has already been
exploited, and
\action\ is supposed to be already renormalized.  Notice also that the
elimination of
$\Psi$ and  $\bar \Psi$ deprived us of the Ito term that is useful
 non-perturbatively~\zbgr.

We next translate $b_\m \to b_\m - i D_\l F_{\l\m}$.  The cross-term
$F_{5\m}D_\l F_{\l\m}$ is an exact derivative because of the Bianchi
identity, and we obtain
\eqn\fact{\eqalign{I = \int d^5x [& ib_\m F_{5\m} + (1/2)b_\m^2
+ (1/2)(D_\l F_{\l\m})^2 \cr
& + il(aA_5 - \p_\m A_\m) + \lb(aD_5 - \p_\m D_\m)\l]
\ .}}
If one identifies $b_\m$ as a 4-component color-``electric" field, this
resembles Faddeev--Popov theory in first-order formalism, with a particular
gauge-fixing term, except that the magnetic energy
$(1/2)B_i^2$ is replaced by $ (1/2)(D_\l F_{\l\m})^2$.  This action still has
$\WW $-invariance, but $s$-invariance is lost.

        As a second step we integrate out the remaining ghost and auxiliary
fields
$\l, \lb, b_\m$, and $b_5$.  Noting that the ghost determinant is again a
constant, we obtain the 5-dimensional non-Abelian gauge action in
second-order formalism
\eqn\pact{\eqalign{I[A] = \int d^5x [ (1/2)(\p_5 A_\m
- a^{-1}D_\m \p_\l A_\l)^2
+ (1/2)(D_\l F_{\l\m})^2 ] \ .}}

Here $A_5$ is gauge fixed to
$A_5 = a^{-1}\p_\l A_\l$, a gauge condition often referred to as {\it
stochastic} gauge-fixing \dan.  Indeed the functional integral associated with
the above 5-dimensional action represents the solution
of a Fokker-Planck or Langevin equation.  The latter describe a stochastic
process that may be simulated numerically, and in which the variable
$t = x_5$  counts sweeps over a 4-dimensional lattice.

        The striking feature of the present approach is that the ghost
determinant is
unity as a result of the parabolic ghost equation.  This is a strong
indication that
the Gribov ambiguity is not a difficulty in the present formulation.   However
we have not directly addressed here the issue of establishing that the
action \pact\ is a valid quantization of the 4-dimensional Yang--Mills theory.
We shall return to this topic on another occasion.

\newsec{Conclusion}

We have reconsidered the quantization of 4-dimensional gauge theories in
5 dimensions with particular attention to the invariance needed to
characterize physical observables.   We implemented this invariance by a
BRST-operator $\WW $, with $\WW ^2 = 0$, that encodes ordinary gauge
invariance, and physical operators $O$ are required to satisfy $\WW {  O} =
0$, and $  { O} \neq \WW (X) $.  The
operator $\WW $ is to be distinguished from the topological BRST-operator $s$,
with $s^2 = 0$, that will occur in the   quantization of all theories with
an additional time,
including gauge and non-gauge theories.  The two operators are compatible
in the sense that $(s + \WW )^2 = 0$.
Together, these two symmetries are extremely restrictive, and we have
constructed the most general action $I$, eqs. \tinv\ and \tgf, that is
invariant
under both symmetries $sI = \WW I = 0$, and that is renormalizability by power
counting.  In Appendix A we derived the Ward identities associated with these
2 symmetries.  In the other Appendices we have examined various aspects and
extensions of the theory such as the Landau-gauge limit, a confinement
scenario, spinor fields, anomalies and lattice discretization without
fermion doubling.

The  esthetics of the formulation in five dimensions, and its attractive
geometric interpretation, strongly suggest that it should be adopted as a
starting
point for defining a gauge theory. As we have seen  in our discussion of
anomalies,  Appendix E, the fifth time acts as a regulator.  Indeed the
theory produces correlation functions in five dimensions, and the physical
limit requires taking a slice in the fifth time.   However new divergences
appear when the times coincide (as is obvious in momentum space
where setting the times equal corresponds to additional integrations over the
conjugate momenta).  This is our  interpretation of the chiral
anomaly.

        The 5-dimensional formulation provides the stage for a
simple confinement scenario in the minimal Landau gauge.  In this scenario, the
long-range confining ``force" is transmitted by $A_5$, the fifth component of
the gluon field.  It is suggestive that the imposition of the analog of Gauss's
law in the fifth dimension can produce a  confinement scenario similar to the
one in the Coulomb gauge \gribov\ and \rcoulomb,
where the long-range confining force is transmitted by $A_4$.  However the
scenario is compatible with manifest 4-dimensional Lorentz invariance.

        We use the $s$ and $\WW $ invariance to improve
the argument that we gave in \zbgr\ to   prove renormalizability of the
action in 5 dimensions.  This requires that we also address possible anomalies.
We exhibit an interesting cocycle in 5 dimensions (solution of the
Wess--Zumino consistency conditions for $s$ and $\WW $ symmetries).
However its coefficient appears to be 0 in general, because of the topological
nature of the action.  Consequently the Noether currents $K_M$ ($M =
1,...,5$) of the 5-dimensional action are strictly conserved.  The origin of
the 4-dimensional anomaly may be understood when one
recognizes that the currents of physical interest are not the $K_M$, but
rather the Noether currents $J_\m$ ($\m = 1,...,4$) of the 4-dimensional
action.  These two different currents are not related in any obvious way.
Indeed the $J_\m$ are not conserved in the 5-dimensional theory in general.
However consistency requires that the $J_\m$ generate the appropriate
4-dimensional Ward identities when inserted into correlation functions at a
fixed fifth time.  This may fail due to singularities that appear in the
correlation functions when the times are set equal.  Indeed we have verified by
explicit computation that the ABBJ triangle anomaly appears as a discontinuity
of the 3-point $J$-current correlator as equal times are approached.  Power
counting in 5-dimensions leads us to expect that there are no other such
breakdowns.

The principles of locality, gauge symmetry and power counting also
determine the form of the topological action for spinors, whether chiral or
not.  It is gratifying that this yields a convergent functional integral
that gives well-defined correlation functions.  It should be emphasized
that the stochastic interpretation of the fifth coordinate must be
abandoned when the local five-dimensional theory is extended to include
fermions.  Remarkably, the 5-dimensional spinor action allows a natural
lattice discretization that does not suffer from fermion doubling, as is
shown in Appendix F.  It is not necessary to introduce a domain wall for this
purpose \kaplan, \neubergerr, \shamir.

In summary, the five-dimensional formulation of renormalizable gauge
theories appears as a very powerful tool for investigating the non
perturbative questions relative to the Yang--Mills theory.

  \vskip .5cm
{\centerline{\bf Acknowledgments}}

The research of Daniel Zwanziger was partially supported by the National
Science Foundation under grant PHY-9900769. P.A. Grassi is supported by
National Science Foundation under grant PHY-9722083.
\vskip .5cm

\appendix A{Ward Identities, Renormalization, and Stability of the Action}

        We shall show that with the action \action\ the combination $g\l$ is
invariant under renormalization,
\eqn\invar{\eqalign{g\l = g_r \l_r  }}
Moreover in the Landau gauge limit $a \to 0$, the quantity $g A_5$ is
invariant also invariant under renormalization.

In \rcoulomb\ and \coulomb\ it was
shown that $gA_4$ is invariant under renormalization in the Coulomb
gauge in the 4-dimensional formulation.  For the same reason $gA_5$ is
invariant under renormalization in 5-dimensional formulation in the Landau
gauge.  [The result depends on the following argument.  The component
$j_5$ of the conserved Noether current of the $\WW $-symmetry is given by
$j_5 = \p \CL / \p (\p_5 A_\m) D_\m \l = ib_\m D_\m \l$, because all
derivatives with respect to $t = x_5$ are contained in
$F_{5,\m} = \p_5 A_\m - D_\m A_5$.  Moreover the left-hand side of
Gauss's law is given by
$\d I / \d A_5 = i D_\m b_\m$.  This allows us to write the conserved
Noether charge $Q = \int d^4x  \ j_5$ as
$Q = - \int d^4x \ \l \d I/ \d A_5$.  No product of fields appears
in this expression, so it provides a Ward identity that is special to this
gauge,
and it leads to the condition
$Z_{A_5}Z_g = 1$  on the renormalization constants
$Z_{A_5}$ and $Z_g$ of $A_5$ and $g$.]


The renormalization of the model contains some interesting features that
we discuss in the present appendix. First, we will show that the rich
content of the
symmetry (the $s$-symmetry and the $\WW $-symmetry \defta) provides
strong constraints on the form of the gauge fixed action; second, we will
show that
the particular gauge choice \tgf ~implies non-renormalization properties
for the ghost fields.

In order to discuss the set of functional equations which implement the
symmetries at the quantum level,
we introduce the sources coupled to the variations of the fields. The
latter are needed to
renormalize theories  with non-linear transformations as in our case. Given
$\phi$,
a  field of the set $\{A_\mu,A_5,\Psi_\mu,\Psi_5,\Phi,c, \bar\Psi_\mu,
b_\mu, \bar\Phi, \bar\eta \}$,
we introduce its ($\WW s$)-source $\phi^*$, its $\WW $-source $\phi'$, and
its $s$-source $\phi''$.
The statistic of $\phi', \phi''$ is opposite to that of $\phi$, and the
statistic of $\phi^*$
is the same of $\phi$. On the
other hand, the ghost numbers of $\phi', \phi'',\phi^*$ are easily fixed by
their couplings. In particular for
a field $\phi$ with ghost numbers $(N_s,N_w)$, we have
\eqn\antifields{\eqalign{
I_{\rm sources}[\phi] &= \int d^5x \sum_\phi \WW  \Big[ s ( \phi^* \phi) \Big]
=\int d^5x  \sum_\phi \WW  \Big[ (s \, \phi^*) \phi  + (-)^{N_s + N_w}
\phi^*  s \, \phi \Big] = \cr
=\sum_\phi  \int & d^5x \Big[  (\WW  s  \, \phi^*)  \phi + (-)^{N_s +
N_w+1} (s \, \phi^*) \WW  \phi +
(-)^{N_s + N_w} (\WW  \, \phi^*) s \,\phi  +  \phi^* \WW  s  \, \phi  \Big]\,.
}}
Adding a $\WW s$-trivial term does not modify either the $s$- or the $\WW
$-cohomology
of the classical theory. However, at the quantum level, we have to
translate the
$s$- and $\WW $-transformations in terms of nilpotent functional operators,
and therefore,
we also require
\eqn\antitrans{\eqalign{
s \phi^* &= (-)^{1+ N_s + N_w} \phi'\,,~~~~
\WW  \phi^* = (-)^{N_s + N_w} \phi'' \,,~~~~
\WW  s \phi^* = - s \WW  \phi^* =0 \,, \cr
s \phi' &= 0\,, ~~~~ \WW  \phi' = 0\,,~~~~
s \phi'' = 0\,, ~~~~ \WW  \phi'' = 0\,.~~~~
}}
Consequently, the eq.~\antifields~becomes
\eqn\sources{\eqalign{
I_{\rm sources}[\phi] = \sum_\phi \int d^5x \Big[  \phi' \, (\WW  \phi) +
\phi'' \, (s \,\phi)  +  \phi^* (\WW  s  \, \phi)  \Big]\,.
}}
As concerns the fields $\lambda,\mu$ and the anti-ghosts
$\bar\lambda,\bar\mu,\bar c$,
due to the fact that their
$s$ and $\WW $-transformations are almost trivial, we can spare several
sources, but we cannot avoid
the following two terms
 \eqn\ghostsources{\eqalign{
I_{\rm ghost} =  \int d^5x \Big[ \lambda' \Big( -{1 \over 2}
[\lambda,\lambda]\Big) - \lambda^* [\lambda,\mu]
\Big]\,,
}} where we identify $\lambda^*$ with $\mu'$. More specifically, we do not
introduce sources for the antighost fields $\bar\lambda,\bar c$ and
$\bar\mu$, and for the fields $\lambda$ and $\mu$, we have $s \lambda^* = -
\lambda'$ and
$\WW  \lambda^* = 0$ which is consistent with the identification
$\lambda^*=\mu'$. In the following, the symbol
$I$ denotes the action including the source terms.

Following the conventional procedure \books\ and given the source terms
\sources~and \ghostsources~, we can establish the
functional identities for $s$- and $\WW $-symmetry for the generating
functional $\Gamma$ of irreducible Green functions
\eqn\STI{\eqalign{
{\cal S}( \Gamma ) &=
\int d^5x \Big\{\sum_\phi \Big[ {\delta  \Gamma \over \delta \phi }{\delta
\Gamma \over \delta \phi''} -
(-)^{N_s+N_w} \phi' {\delta  \Gamma \over \delta \phi^*}  \Big] + \mu
{\delta  \Gamma \over \delta \lambda}
- \lambda' {\delta  \Gamma \over \delta \lambda^*}  +
 \bar\lambda {\delta  \Gamma \over \delta \bar \mu} -
l {\delta  \Gamma \over \delta \bar c} \Big\}= 0\,,     \cr
{\cal W}( \Gamma ) &=
\int d^5x \Big\{ \sum_\phi \Big[ {\delta  \Gamma \over \delta \phi }{\delta
\Gamma \over \delta \phi'} -
(-)^{N_s+N_w} \phi'' {\delta  \Gamma \over \delta \phi^*}\Big] +
 {\delta  \Gamma \over \delta \lambda }{\delta  \Gamma \over \delta
\lambda'}  +
{\delta  \Gamma \over \delta \mu }{\delta  \Gamma \over \delta \mu'}  +
\bar c {\delta  \Gamma \over \delta \bar\mu} +
l {\delta  \Gamma \over \delta \bar\lambda}
\Big\}= 0\,,
}}
where the sum is extended to the entire set of fields $\phi$ of the present
model. The nilpotency and
the commutation relations between the $s$-symmetry and the $\WW $-symmetry
are expressed by the
following equations
\eqn\commu{\eqalign{
{\cal S}^2_\Gamma =0\,,~~~~~ {\cal W}^2_\Gamma =0\,, ~~~~~
\Big\{  {\cal W}_\Gamma , {\cal S}_\Gamma \Big\}=0\,,
}}
where ${\cal S}_\Gamma$ and  ${\cal W}_\Gamma$ are the linearized version
of the functional operators
involved in \STI~and are defined by ${\cal S}_\Gamma (\Xi) \equiv {\partial
\over \partial \epsilon}
\left. {\cal S}( \Gamma +\epsilon \Xi )\right|_0$. Besides eqs.~\STI, one
has to take into account the equations of motion for the
Lagrangian multiplier $l$ (cf. eq.~\tgf )
\eqn\gauge{\eqalign{
{\delta \Gamma \over \delta l} = a A_5 - \partial_\mu A^\mu\,,
}}
where $a$ is a gauge parameter. From the five-dimensional point of view,
the gauge fixing  \gauge~appears like
a Landau gauge fixing type, but from the four-dimensional point of view,
$A_5$ plays the r\^ole of the Lagrangian multiplier and
the gauge fixing \gauge~is a truly Lorentz gauge type. In the following we
will show that, for a generic $a$ the theory satisfies a set
of functional equations for the ghost fields $\lambda,c,\mu$. Those
equations imply the non-renormalization of the ghost fields themselves
and, in the case  $a= 0$, a further equation implies the
non-renormalization of the combination $ g A_5$ field.

The existence of a solution to the system of equations \gauge,~\STI~is
based on the Fr\"obenius theorem whose main
hypothesis is the existence of an algebra of vector fields on the manifold
spanned by functionals like $\Gamma$. The algebra is
constructed by computing the commutation relations between the
eqs~\STI~(cf. \commu) and with the eq.~\gauge.
For a generic functional ${\cal F}$, they read
\eqn\ghosteq{\eqalign{
{\cal W}_{\cal F} \Big( {\delta{\cal F}\over \delta l} - a A_5 +
\partial_\mu A^\mu
\Big) - {\delta \over \delta l} {\cal W}( {\cal F})  &= {\delta {\cal F}
\over \delta \bar\lambda} - a  {\delta {\cal F}\over \delta A'_5} +
\partial_\mu {\delta {\cal F}\over \delta A'_\mu}\,,  \cr
{\cal S}_{\cal F} \Big( {\delta{\cal F}\over \delta l} - a A_5 +
\partial_\mu A^\mu
\Big) - {\delta \over \delta l} {\cal S}( {\cal F})  &= {\delta {\cal F}
\over \delta \bar c} - a  {\delta {\cal F}\over \delta A''_5} +
\partial_\mu {\delta {\cal F}\over \delta A''_\mu} \,,
}}
and, by commuting again the resulting equations with the operators ${\cal
S}_{\cal F},{\cal W}_{\cal F}$, we have
\eqn\ghosteqs{\eqalign{
{\cal W}_{\cal F} \Big(  {\delta {\cal F} \over \delta \bar c} - a  {\delta
{\cal F}\over \delta A''_5} +
\partial_\mu {\delta {\cal F}\over \delta A''_\mu}
\Big) -  \Big( {\delta \over \delta \bar c} - a  {\delta \over \delta A''_5} +
\partial_\mu {\delta \over \delta A''_\mu} \Big) {\cal W}( {\cal F})  &=
 {\delta {\cal F} \over \delta \bar\mu} - a  {\delta {\cal F}\over \delta
A^*_5} +
\partial_\mu {\delta {\cal F}\over \delta A^*_\mu}  \,, \cr
{\cal S}_{\cal F} \Big(  {\delta {\cal F} \over \delta \bar\lambda} - a
{\delta {\cal F}\over \delta A'_5} +
\partial_\mu {\delta {\cal F}\over \delta A'_\mu}
\Big) -  \Big( {\delta \over \delta \bar\lambda} - a  {\delta \over \delta
A'_5} +
\partial_\mu {\delta \over \delta A'_\mu} \Big) {\cal S}( {\cal F})  &=
 {\delta {\cal F} \over \delta \bar\mu} - a  {\delta {\cal F}\over \delta
A^*_5} +
\partial_\mu {\delta {\cal F}\over \delta A^*_\mu}  \,.
}}
All the other commutation relations are trivial and therefore
\STI,~\ghosteq,~\ghosteqs~describe all the possible non-vanishing
commutation relations and the algebra is, in fact, closed. We notice
that the commutation relations between the ghost equations \ghosteq~with
the operators ${\cal S}_{\cal F},{\cal W}_{\cal F}$ generate the
same functional equation for the ghost $\mu$. This fact is a
consequence of the anti-commutation relations between ${\cal S}_{\cal F}$
and ${\cal W}_{\cal F}$.  Finally, the
equations~\ghosteq,~\ghosteqs~expressed in term of $\Gamma$ can be
easily integrated. This amounts to a redefinition of the sources
$A'_M,A''_M$, and $A^*_M$. We will not discuss these details since
they are common to the conventional procedure of the BRST quantization
with linear gauge fixings.

More interesting and specific to the present model are further
functional equations for the antighost fields $\bar\lambda, \bar c$
and $\bar \mu$. These can be derived by analyzing the corresponding
tree level equations and observing that the composite operators which
appear in those formulae are already present in the extended action
$I$. From eqs.~\gfcaa,~\ccaa,~\caa,~and \sources, we have
\eqn\antighotree{\eqalign{
{\delta I \over \delta \lambda(x)} & =
\Big(a D_5 \bar \lambda - D_\mu \partial^\mu  \bar \lambda \Big) - a \Big[
\Psi_5+  D_5 c, \bar\mu \Big]
- \Big[ \Psi_\mu+  D_\mu c, \partial^\mu \bar\mu \Big] + [\lambda',\lambda]
+ [\lambda^*,\mu] \cr
&+ \sum_\phi \int d^5y \Big[ (-)^{1+N_s + N_w} \phi'  {\delta \over \delta
\lambda(x)} (\WW  \phi)(y) +
(-)^{N_s + N_w } \phi^*  {\delta \over \delta \lambda(x)} (\WW  s
\phi)(y)\Big] \,, \cr
{\delta I \over \delta c(x)} & =
\Big(a D_5 \bar c - D_\mu \partial^\mu \bar c \Big)  + a D_5 \Big[ \bar
\mu, \lambda \Big]
- D_\mu \Big[ \partial^\mu \bar \mu, \lambda \Big] +
[c',\lambda] \cr
& + \sum_\phi \int d^5y \Big[ (-)^{N_s + N_w+1} \phi''  {\delta \over
\delta c(x)} (s \phi)(y) +
(-)^{N_s + N_w } \phi^*  {\delta \over \delta c(x)} (\WW  s \phi)(y)  \cr
& + (-)^{N_s+N_w+1} \phi'  {\delta \over \delta c(x)} (\WW  \phi)(y)
\Big]\,, \cr
{\delta I \over \delta \mu(x)} & =
\Big(a D_5 \bar\mu - D_\mu \partial^\mu  \bar \mu \Big)  - [\lambda^*,
\lambda] - c' +
\sum_\phi \int d^5y \Big[ \phi^*  {\delta \over \delta \mu(x)} (\WW  s
\phi)(y) \Big]\,.
}}
It is important to note that the terms proportional to the sources
$\phi',\phi''$ are linear in the quantum fields and therefore they do
not require an independent renormalization besides the usual field
renormalization. As an example, some of those terms are explicitly
shown in the last equation. On the other hand, the last terms
proportional to the sources $\phi^*$ are not linear in quantum fields
and therefore they require more care. Eqs.~\antighotree~are not
suitable for quantization since they involve the renormalization of
operators like $\Big(a D_5 \bar \lambda - D_\mu \partial^\mu \bar \lambda
\Big)$
which, unfortunately, do not belong to the set of those
coupled to the sources~\sources~and~\ghostsources. However,
considering the integrated (over the 5-dimensional manifold) version
of eqs.~\antighotree~ and integrating by parts, we have the following
equations for $\Gamma$:
\eqn\antighost{\eqalign{
\int d^5x {\delta \Gamma \over \delta \lambda(x)} =
\int d^5x &\Big\{ \Big[\bar \lambda, {\delta \Gamma \over \delta l}\Big] +
\Big[\bar \mu, {\delta \Gamma \over \delta \bar c}\Big]
+ [\mu',\mu]+ [\lambda',\lambda] \cr
& +  \sum_\phi   [\phi' , \phi] +
\Big[ \phi^*,  {\delta \Gamma \over \delta \phi''} \Big] \Big\}\,,
\cr
\int d^5x {\delta \Gamma \over \delta c(x)}  =
\int d^5x &\Big\{ \Big[\bar c, {\delta \Gamma \over \delta l}\Big] +
\Big[\bar \mu, {\delta \Gamma \over \delta \bar \lambda}\Big]
 + \Big[\lambda, {\delta \Gamma \over \delta \mu}\Big] + [c',\lambda] \cr
&+  \sum_\phi [\phi'', \phi]  +
\Big[ \phi^* , {\delta \Gamma \over \delta \phi'} \Big] \Big\}\,,
\cr
\int d^5x {\delta \Gamma \over \delta \mu(x)} =
\int d^5x &\Big\{ \Big[\bar \mu, {\delta \Gamma \over \delta l} \Big]
 - [\lambda^*, \lambda] - c' +  \sum_\phi (-)^{N_s+N_w}\Big[ \phi, \phi^*
\Big]\Big\}\,.
}}
Notice that, by using the sources $\phi'$ and $\phi''$, we are able to
translate from the tree level approximation \antighotree~ to the quantum
level also the terms coming from the $s\WW $ variations.  As a check of
the procedure, we can compute the anti-commutation relations between
the anti-ghost equations \antighost~and the ghost equations
\ghosteq~and \ghosteqs, to see that they close on the gauge-fixing
equation \gauge~and they are compatible with each other. This supports
the hypothesis of the Fr\"obenius theorem which allows us to integrate
the complete system of equations. In particular, it is important to
compute the commutation relation between the ${\cal W}_\Gamma$ and the
first equation of \antighost~ or that of ${\cal S}_\Gamma$ with the
first one. This gives a new functional equation
\eqn\WTI{\eqalign{
\int d^5x & \Big\{
\Big[\bar \lambda, {\delta \Gamma \over \delta \bar \lambda}\Big] +
\Big[\bar \mu, {\delta \Gamma \over \delta \bar \mu}\Big] +
\Big[\bar c, {\delta \Gamma \over \delta \bar c}\Big] +
\Big[\lambda, {\delta \Gamma \over \delta \lambda}\Big] +
\Big[\mu, {\delta \Gamma \over \delta  \mu}\Big]  \cr
&
+ \sum_\phi
\Big[  \phi , {\delta \Gamma \over \delta \phi} \Big] +
\Big[  \phi' , {\delta \Gamma \over \delta \phi'} \Big] +
\Big[  \phi'' , {\delta \Gamma \over \delta \phi''} \Big] +
\Big[  \phi^* , {\delta \Gamma \over \delta \phi^*} \Big]
+ [c',\mu] \Big\} =0\,,
}}
which implements the invariance under the rigid gauge transformations of
the model. As in the case of Yang--Mills, quantized
with the conventional BRST technique, in the Landau gauge, rigid gauge
invariance is a by-product of the
dynamics of the ghost fields \bla.

Equations \antighost~ control the renormalization of the
ghost fields $\lambda, c, \mu$ and, in particular, by means of those
equations the ghost fields have
no independent renormalization. This is a common feature of Landau type
gauge fixing \bla.

To derive these non-renormalization properties, and for pedagogical
purposes, we consider a model
without chiral fermions and we assume an
invariant regularization scheme. In that framework, we can renormalize the
model multiplicatively. That is, all the fields will
be renormalized by a suitable wave function renormalization $\phi
\rightarrow Z_\phi\phi$.
Requiring that eqs. \gauge,~\ghosteq,~\ghosteqs,~and \antighost~ are preserved
by the renormalization procedure, we immediately get
 \eqn\antizeta{\eqalign{
Z_a &= Z_A Z_5^{-1}\,,~~~~
Z_{\bar\mu} = Z_{\bar c} = Z_{\bar \lambda} = Z_l = Z^{-1}_A\,, \cr
Z_\mu &= Z_l Z^{-1}_{\bar\mu} = 1\,,~~~~
Z_\lambda = Z_l Z^{-1}_{\bar\lambda} = Z^{-1}_{\bar\mu} Z_{\bar c}=1\,, \cr
Z_c &= Z_l Z^{-1}_{\bar c} = Z^{-1}_{\bar\mu} Z_{\bar \lambda} = Z_{\mu}
Z^{-1}_{\lambda} =1\,,
}}
where we used the rescaled gauge fields $A_M \rightarrow g\, A_M$. Due to
the relations
$Z_\lambda = Z_c = Z_\mu = 1$, it is direct to conclude that the products $
g\, \lambda, g\, c$ and $g\, \mu$
do not renormalize.

Finally we can switch off the gauge parameter $a$ by letting it go to zero.
In this case, the analysis can be
repeated obtaining the same results for the non-renormalization of the
ghost fields. However, instead of
integrating the anti-ghost equations \antighotree~over the five-dimensional
space,
we can also integrate only over the four-dimensional
Lorentz invariant manifold.
This implies that the anti-ghost equations \antighost, rewritten with
integrals over four dimensions, are
\eqn\newantighost{\eqalign{
\int d^4x {\delta \Gamma \over \delta \lambda(x)} =
\int d^4x & \Big\{\partial_5 A'_5 + \Big[\bar \lambda, {\delta \Gamma \over
\delta l}\Big] +
\Big[\bar \mu, {\delta \Gamma \over \delta \bar c}\Big]
+ [\mu',\mu]+ [\lambda',\lambda] \cr
& + \sum_\phi   [\phi' , \phi] +
\Big[ \phi^*,  {\delta \Gamma \over \delta \phi''} \Big] \Big\}\,,
\cr
\int d^4x {\delta \Gamma \over \delta c(x)} =
\int d^4x & \Big\{\partial_5 A''_5 + \Big[\bar c, {\delta \Gamma \over
\delta l}\Big] +
\Big[\bar \mu, {\delta \Gamma \over \delta \bar \lambda}\Big]
 + \Big[\lambda, {\delta \Gamma \over \delta \mu}\Big] + [c',\lambda] \cr
&+  \sum_\phi [\phi'', \phi]  +
\Big[ \phi^* , {\delta \Gamma \over \delta \phi'} \Big] \Big\}\,,
\cr
\int d^4x {\delta \Gamma \over \delta \mu(x)} =
\int d^4x & \Big\{-\partial_5 A^*_5 + \Big[\bar \mu, {\delta \Gamma \over
\delta l} \Big]
 - [\lambda^*, \lambda] - c' +  \sum_\phi (-)^{N_s+N_w}\Big[ \phi, \phi^*
\Big]\Big\}\,.
}}
They are local in the fifth component of the 5-dimensional space and
satisfy all the commutation relations
among themselves and with the other functional operators with the proper
obvious modifications. However,
due to the presence of new terms like $\partial_5 \int d^4x A'_5$,
not killed by the integration over the 5-dimensional space, the commutation
relation of the first
equation of \newantighost~with
${\cal W}_\Gamma$, or the second one with ${\cal S}_\Gamma$, generates the
new functional equation
\eqn\localWTI{\eqalign{
- \, \partial_5 & \int d^4x {\delta \Gamma \over \delta A_5} +
\int d^4x  \Big\{
\Big[\bar \lambda, {\delta \Gamma \over \delta \bar \lambda}\Big] +
\Big[\bar \mu, {\delta \Gamma \over \delta \bar \mu}\Big] +
\Big[\bar c, {\delta \Gamma \over \delta \bar c}\Big] +
\Big[\lambda, {\delta \Gamma \over \delta \lambda}\Big] +
\Big[\mu, {\delta \Gamma \over \delta  \mu}\Big]  \cr
&
+ \sum_\phi
\Big[  \phi , {\delta \Gamma \over \delta \phi} \Big] +
\Big[  \phi' , {\delta \Gamma \over \delta \phi'} \Big] +
\Big[  \phi'' , {\delta \Gamma \over \delta \phi''} \Big] +
\Big[  \phi^* , {\delta \Gamma \over \delta \phi^*} \Big]
+ [c',\mu] \Big\} =0\,.
}}
This equation implements the invariance under gauge transformations, local in
the fifth  component, and,  automatically, implies that the fifth component of
the gauge field $A_5$ is not renormalized, or equivalently,  by rescaling
properly the gauge fields by the gauge coupling $g$, the combination $g A_5$
is not renormalized.  We may use the $gA_5$ propagator to define an
invariant charge in QCD.   A similar situation happens for the abelian sector

of the Standard Model quantized in the background gauge \grassi.

To conclude the present section, we would like to stress that for a generic $a$
the content of the symmetry is so rich that it constrains the  form of the
action
strongly. Indeed, by a tedious algebra, one can easily prove that the only
solution of the system which satisfies the  power-counting constraints is
given by
the action $\tinv$.

 \appendix B{Landau Gauge Limit}

We wish to consider the important limiting
case in which the gauge parameter $a \to 0$, and the 4-vector potential $A_\m$
becomes transverse
$\p_\m A_\m = 0$.  Like the Coulomb gauge condition $\p_i A_i = 0$ in the
4-dimensional formulation, this gauge condition is a singular in the present
5-dimensional formulation because it leaves $t$-dependent but
$x$-independent gauge-transformations $g(t)$ unfixed (cf. eq.~\localWTI).
However it is
expected that the limit $a \to 0$ is finite in the sense that the renormalized
correlation functions calculated at positive $a$ have a finite limit as $a \to
0$.  Indeed this property has been established in the 4-dimensional
formalism when the Coulomb gauge is approached from an interpolating
gauge \coulomb.  Without attempting here to establish the finiteness of the
limit $a \to 0$ in the present case, we shall instead derive some of its
properties.

        The transversality condition $\p_\m A_\m = 0$ is generally called the
Landau gauge.  However this is not a well-defined gauge because of
the Gribov ambiguity, and there is really an infinite class of Landau gauges.
We shall show that in the limit $a \to 0$ we end up in a gauge, which we call
the ``minimal" Landau gauge, in which the additional condition
\eqn\greg{\eqalign{M(A) \equiv - \p_\m D_\m(A) \geq 0 \ ,}}
is satisfied.  This states that the Faddeev--Popov operator $M(A)$ is
non-negative.  It is a condition on $A_\m$ that defines the (first) Gribov
region.   By contrast, the Faddeev--Popov method in 4 dimensions does not
restrict the functional integral in the Landau gauge to the Gribov region.
Indeed, if one attributes a non-perturbative significance to the
Faddeev--Popov formula by BRST quantization, then a signed sum over the
entire region both inside and outside the Gribov horizon is implied.

        To study the limit $a \to 0$, we rescale $t \to a t$ in \pact\ and
obtain
\eqn\sact{\eqalign{I = \int d^5x [ (2a)^{-1}(\p_5 A_\m - D_\m \p \cdot A)^2
+ (a/2)(D_\l F_{\l\m})^2 ] \ ,}}
where $\p \cdot A  \equiv \p_\m A_\m$.  As $a$ approaches 0, the first term
dominates and the probability gets concentrated at its absolute minimum
namely at
\eqn\flg{\eqalign{\p_5 A_\m = D_\m \p \cdot A \ .}}
With $t = x_5$ as the time variable, this equation
defines a flow that is tangent to the gauge orbit, with infinitesimal
generator
$\omega = \p \cdot A$.  The {\it global},
non-perturbative character of this flow will be deduced from the fact that it
describes steepest descent of the functional
\eqn\minf{\eqalign{F = (1/2)\int d^4x \ A_\m^2}}
restricted to the gauge orbit.  For under an arbitrary infinitesimal gauge
transformation  $\d A_\m = D_\m \omega$, the functional undergoes the
infinitesimal variation
\eqn\desc{\eqalign{\d F  =  \int d^4x A_\m  \d A_\m
=  \int d^4x A_\m D_\m \omega = - \int d^4x \ \p_\m A_\m \ \omega  ,
}}
so steepest descent of this functional restricted to the gauge orbit is indeed
achieved with $\omega = \p_\m A_\m$.  Since the functional is bounded
below, the steepest descent necessarily approaches a local
minimum.\foot{This argument is rigorous in the 5-dimensional compact
lattice-gauge formulation \zbgr.  The classical descent could in principle end
at a saddle-point but quantum fluctuations prevent this outcome.} At a
minimum, the functional is stationary, which yields the Landau gauge
condition  $\p \cdot A = 0$.  Moreover its second variation is non-negative at
a minimum, which, from \desc, gives the additional condition,
\eqn\hor{\eqalign{ \int d^4x \ \omega \ (- \p \cdot D) \omega \geq
0 \   {\rm for \ all} \ \omega \ ,}}
which establishes \greg.  Thus in the limit
$a \to 0$,  the gauge-fixed functional integral \sact, which is local in 5
dimensions, has the 4-dimensional property of concentrating the
probability  within the Gribov horizon.\foot{Because
there are in general more than one relative minimum on a gauge orbit, the
restriction to the Gribov region \greg\ is not a complete gauge fixing.
However the validity of the present formulation does not in any way depend
on how the probability may be distributed among the various possible relative
minima in the limit $a \to 0$, nor, for that matter, does it depend on taking
the limit $a \to 0$ at all, but rather is valid for any value of $a \geq 0 $.}

        Remarkably, the Landau gauge limit just obtained from the {\it
local} 5-dimensional action coincides with the ``minimal" Landau gauge that
is used in numerical studies of lattice gauge theory.  Indeed in these studies,
the gauge is fixed by numerically minimizing a lattice analog of the
functional $F = (1/2)\int d^4x A_\m^2$ restricted to the gauge orbit, so that
the configuration is likewise brought to a local minimum of the functional,
namely to a point within the first Gribov region.  Thus the 5-dimensional
formulation provides a BRST-invariant, renormalizable, continuum
description of the minimal Landau gauge that is accessible numerically.
Moreover other numerically accessible gauges such as the minimal Coulomb
gauge or maximal abelian gauge may be introduced in the 5-dimensional
formalism by imposing the 5-dimensional gauge condition $aA_5 = - G(x)
F$, where $F$ is an appropriately chosen functional, and
$G(x) = - D_\m \d F/\d A_\m$ is the gradient in the gauge orbit direction.
These are not in the class of Faddeev--Popov gauges in which at the
non-perturbative level a signed sum over the entire region outside
the Gribov region is implied.  Needless to say, if one wishes to compare
numerically gauge-fixed quantities such as propagators with analytic
predictions such as, for example, the Nielsen identities~\nie, they should be
derived in the same gauge.

\appendix C{Confinement Scenario}

        The present 5-dimensional formulation provides a simple confinement
scenario in the Landau gauge limit.  The basic idea is that $gA_5$, which is
a Lorentz scalar, provides a vehicle for the transmission of long-range
correlations that correspond to a confining force.  In this respect it
resembles
the component $gA_4$ of the gluon field in the Coulomb gauge.  In fact the
scenario was originally developed for the Coulomb gauge by Gribov \gribov\
and elaborated more recently \rcoulomb.  However in the present
formulation, the mechanism respects manifest 4-dimensional Lorentz
invariance.  The discussion that follows is
non-perturbative, and could be based on the lattice-gauge formulation of the
5-dimensional formalism presented in \zbgr.

        It is shown in Appendix A that in the Landau gauge limit, the
field $gA_5$
is invariant under renormalization
\eqn\invarl{\eqalign{g A_5 = g_r A_{5r} , \
\ {\rm for} \ \ a = 0. }}
This implies that in the Landau gauge limit all correlation functions of
$gA_5$ are renormalization-group invariants.  As such they are finite and
independent of the ultra-violet cut-off, and depend only on the QCD scale
$\L_{QCD}$.  Thus the statement that they are long range refers to the
QCD scale.

        To derive the properties of the Landau-gauge limit, we return to the
first-order action \fact, and integrate out the ghosts $\l$ and $\lb$, which
again gives a constant determinant.  We pose
$a = 0$, and obtain
\eqn\cact{\eqalign{I = \int d^5x [ ib_\m F_{5\m} +
(1/2)b_\m^2 + (1/2)(D_\l F_{\l\m})^2 - il \p_\m A_\m ]
\ .}}
Apart from the substitution $B_i^2 \to (D_\l F_{\l\m})^2$, this is the
Coulomb-gauge action in 5 dimensional space-time, and we shall solve the
constraints just as is in the Coulomb gauge. We integrate out the Lagrange
multiplier field $l$, which gives
$\d(\p_\m A_\m)$, and imposes the gauge constraint $A_\m = A_\m^\tr$.
Integration on $A_5$ which appears in the
action only in the term
$b_\m F_{5\m} = b_\m(\p_5 A_\m - D_\m A_5)$  gives
$\d (D_\m b_\m)$  which imposes a form of Gauss's
law,
\eqn\Gauss{\eqalign{D_\m b_\m = 0}}
on the 4-dimensional color-electric field $b_\m$.  Indeed if we write this as
\eqn\rGauss{\eqalign{\p_\m b_\m =  \r \ ,}}
the color-charge density
$\r = j_5 = - [A_\m, b_\m]$  is the $j_5$-component of the conserved
Noether current $(j_\m, j_5)$ of global gauge invariance.  To solve Gauss's
law, we write
$b_\m = b_\m^\tr - \p_\m \f$, where $\f$ plays the role of a Coulomb
potential for the 4-dimensional color-electric field $b_\m$.  Gauss's
law is solved by
\eqn\solGaus{\eqalign{\f = M^{-1} \r_l \ ,}}
where $\r_l \equiv - [A_\m^\tr, b_\m^\tr]$ is the color-charge
density of the transverse gluons.  Here
$M=M(A^\tr) = - D_\m(A^\tr) \p_\m$, with $\m = 1,...4$,  is the
4-dimensional Faddeev--Popov operator.  Its inverse $M^{-1}(A^{\rm tr})$
is an integral operator that acts {\it instantaneously} in the 5-dimensional
space-time and transmits the long-range force that is expected to be
confining.

        When this expression for $\f$ is substituted back into the action,
it assumes the canonical form\foot{One might expect that the integral over
$\f$ will produce the inverse of the Faddeev--Popov determinant
$\det^{-1}(-\p_\m D_\m)$ which, with $\m = 1,...4$, is ill-defined
in a 5-dimensional functional integral.  However the integral over $\f$
should be performed before setting gauge parameter $a = 0$, which gives
instead $\det^{-1}(a D_5 - \p_\m D_\m)$.  This is a constant for the same
reason that the ghost determinant is a constant, namely the propagator of the
{\it parabolic} operator $a \p_5 - \p_\m^2$ is retarded, so all closed loops
vanish.}
\eqn\reaction{\eqalign{I(A^\tr, b^\tr) = \int dt [
ib_\m^\tr \p_5 A_\m^\tr + H(A^\tr, b^\tr)]
\ ,}}
with ``hamiltonian"
\eqn\ham{\eqalign{H = &  \int d^4x [(1/2)(b_\m^\tr)^2 +
(1/2)(D_\l F_{\l\m})^2 ]   \cr
 & + (1/2)\int d^4x d^4y \ \r_l(x) \ [M^{-1} (- \p^2)M^{-1}](x,y) \ \r_l(y)
\ .}}
This is the Coulomb hamiltonian in one extra space dimension.  The
last term represents the instantaneous color-Coulomb interaction of
separated color charge $\r_l$.

        The preceding derivation may also be used to show that the propagator
\eqn\prop{\eqalign{D_{55}(x,t) = \langle gA_5(x,t) gA_5(0,0) \rangle \ .}}
develops an instantaneous part in the Landau gauge limit,
\eqn\inst{\eqalign{D_{55}(x,t) = V(x) \d(t) + {\rm (non-instantaneous)} .}}
Here $V(x)$ is an analog of the instantaneous color Coulomb
potential in one extra dimension that is given by
\eqn\cpot{\eqalign{V(x-y)
= \langle [M^{-1} (- \p^2)M^{-1}](x,y) \rangle
\ .}}

As shown in Appendix A, the correlation functions of $gA_5$ are
renormalization-group invariants including in particular the instantaneous
part $V(x)$, and its fourier transform ${\tilde V}(k)$.
It is thus of the form ${\tilde V}(k) = g^2(k/\L_{QCD})/k^2$,
where $g(k/\L_{QCD})$ is a running coupling constant that depends only
the QCD mass scale.

        Thus the Landau gauge limit of the 5-dimensional formulation
exhibits all
the features of the Coulomb gauge in one extra space dimension, with the
additional advantage of maintaining manifest 4-dimensional Lorentz
invariance.  Consequently the arguments for confinement in the Coulomb
gauge \gribov, \rcoulomb, may be taken over wholesale to the present case.
Without reproducing these arguments here we recall that the restriction to
the first Gribov region $M(A^{\rm tr}) \geq 0$,  demonstrated in
the previous section, produces 2 related effects. (i)  Configurations
corresponding to small eigenvalues of
$M(A^{\rm tr})$ are favored by entropy.  This makes the Green function
$[M^{-1}(A^{\rm tr})](x-y)$  long range, and consequently also the
instantaneous color-Coulomb potential in \ham and \cpot.  (ii)  The
low frequency components of
$A_\m^{\rm tr}$ are suppressed, thereby eliminating the physical gluons
from the physical spectrum.  Indeed it has been proven in lattice gauge
theory in the Landau or Coulomb gauge that at infinite lattice volume the
gluon propagator
$D_{\m\n}(k)$ of transverse gluons vanishes for any probability distribution
that is restricted to the first Gribov region \dzvan.  It would require
detailed dynamical arguments to determine whether the instantaneous
color-Coulomb potential is sufficiently long range to confine all colored
objects.  However the mechanism of a long-range force that couples
universally to color charge \ham, is clearly present.

\appendix D  {Spinor action and propagator}

The five-dimensional formalism can accommodate the coupling of the
gauge field to spinors, and moreover we observed heuristically in
\zbgr\ that this extended point of view  naturally introduces many of
the ingredients that look rather ad hoc  in the genuine
four-dimensional formulation when one tries to answer
 non-perturbative questions.  We wish to address these issues in more detail,
particularly the question of spinors and  chirality.

        The action that defines the  propagators and the interactions of the
four-dimensional spinors  in the five-dimensional formalism was
introduced in \zbgr.  Our postulate is that one must extend the
definition of the topological BRST $s$-operator to spinors
 (in a way that does not necessitates the introduction of a Langevin
equation with a fermionic noise.)  Naturally this
must be done in a way which is consistent with the symmetries such as the
gauge invariance implemented by $\WW $-symmetry, the chirality, the
conservation of the fermionic number, etc.  Once this is done, the principle
of locality and renormalizability must determine  the form of a
renormalizable local action in five dimensions.  As we will see
shortly, this simple five-dimensional point of view turns out to
be surprisingly predictive.

  Let $q(x)$ be a spinor in four dimensions.  For concreteness we
suppose it describes a quark, with spinor and color indices
suppressed as usual.  Extend it to a five-dimensional object
$q(x) \to q(x,t)$, without affecting the spinorial index.  In accordance
with our postulate, $q$ is a member of a spinor quartet, made up of $q, \P_q,
\bar\P_q$, and $b_q$, which are, besides the anti-commuting spinor $q$, its
commuting topological ghost and anti-ghost, and $b_q$, the anti-commuting
Lagrange multiplier. They each carry unit quark charge, and $s$ and $\WW $ act
on them according to
\eqn\spinor{\eqalign{
(s+\WW )q+(c+\l)q =  \P_q   ~~~~~~~(s+\WW )\P_q+(c+\l)\P_q=  \F\ q ~\cr
(s+\WW )\bar\P_q + (c+\l)\bar\P_q=  b_q
~~~~~~~(s+\WW )\b_q + (c+\l) b_q =  \F \ \bar\P_q,
}}
where $c, \l$ and $\F$ act on the spinors in (say) the fundamental
representation, $c \equiv c^a t^a$, where $t^a$ are the Gell-Mann
matrices.  By separate conservation of $N_s$ and $N_w$, this gives
$sq = \P_q - cq$, $s\P_q = - c \P_q - \F q$ and
$s\bar \P_q = - c\bar \P_q +b_q $,
 $sb_q = - c, b_q - \F,\bar \P_q$, and  the $\WW $-symmetry   acts on
the field as an infinitesimal gauge transformation with a parameter
equal to the ghost $\l$.  There is a corresponding independent quartet for the
anti-quark which consists of $q^{\dag},\P_q^{\dag}, \bar\P_q^{\dag}$, and
$b_q^{\dag}$, each with quark charge $-1$.  \foot{Since we already use
``bar" notation to designate an anti-ghost, such as $\Pb$, we
unconventionally use ``dagger" notation, such as $q^\dagger$ instead of
$\bar{q}$, to designate Dirac conjugation.}

 Five-dimensional Lorentz covariance of spinors $q,\P_q,\bar\P_q,b_q$
does not concern us since we only have in view an $SO(4)\times
R$-invariant theory rather than an $SO(5)$-invariant one, namely the
5-dimensional formulation of the standard Dirac action,
\eqn\dirac{\eqalign{  S = \int d^4 x \
q^{\dag} (\g_\m D_\m -m)q.}}  The method of stochastic quantization is
not of direct relevance for the case of anticommuting
fields. Therefore, to determine the 5-dimensional action for spinors,
we postulate that it must be invariant under the topological BRST
symmetry  \spinor. Using power counting and locality requirements,
this must give:

\eqn\diracf{\eqalign{  I_q = \int d^5x \ s\{
\Pb_\m {{\d S} \over {\d A_\m}} \ + \
& \Pb_q^{\dag}[D_5 q - K({{\d S} \over {\d q^{\dag}}} - \demi b_q) ]\cr
& + [ - q^{\dag}D_5
- ({{\d S} \over {\d q}} + \demi b_q^{\dag}) K]\Pb_q  \} \ , }}
that is,
\eqn\diracg{\eqalign{ I_q = \int d^5x \
s \big( \Pb_\m  q^{\dag} \g_\m q \  +  \
&\Pb_q^{\dag}\{ \ [ D_5 - K(\g_\m D_\m - m) ]q
    + { {K}\over{2} } b_q \  \} \cr
& + \{ \ q^{\dag} [ - D_5
   - ( \g_\m D_\m - m)K  ]
   - b_q^{\dag} { {K }\over{2} } \ \} \Pb_q \big).
 }}
The first term contains the anti-ghost $\Pb_\m$ of $A_\m$, and is an
additional contribution to the gluon action \tinv.

 Here is $K$  is a kernel that in principle is at our disposal, because the
expectation value of observables calculated on a given time-slice at
equilibrium is independent of $K$ as can be seen by Ward identities.
However we must choose it to obtain a well-defined functional integral.  In
particular one easily sees that it is necessary that the combination
$K( - \g_\m D_\m + m)$ be bounded below.  If we give the
standard canonical dimension $3/2$ to $q$ and its ghost $\P_q$, and
canonical dimension $5/2$ to the anti-ghost $\Pb_q$ and Lagrange
multiplier $b_q$, and similarly for the
$q^{\dag}$ quartet, and impose gauge-invariance in the form of
$\WW $-invariance, then power counting, $SO(4)$-invariance and gauge
invariance  imply that
\eqn\sol{\eqalign{
K= \g^\m D_\m +M.}}  This is the most general
action which one may write down with these properties, to within
renormalization constants, apart from the fact that the same kernel $K$
 multiplies both $\g_\m D_\m$ and $b_q$.  As a matter of convenience we
shall set the mass
$m$ in the kernel equal to the mass of the Dirac spinor, $M = m$ so the
combination
\eqn\herm{\eqalign{ Q \equiv K( - \g_\m D_\m + m) = KK^\dagger =
K^\dagger K
    = ( - \g_\m D_\m + m)K  = - (\g_\m D_\m)^2 + m^2
 }}
is hermitian and positive and chirally even.  (For well-defined
perturbation theory, $mM \geq 0$ is sufficient.)

                We may consider Weyl spinors in four dimensions, with chiral
coupling  to gauge  fields. The signal that we deal with Weyl spinors is
minimally contained in the range of values over which the spinor index runs,
together with the various $V\pm A$ coupling that may arise in the
interactions. The four-dimensional matrix
$\g^5$ extends trivially  to the fifth $\g$ matrix in five dimensions.

        The extension  to chiral gauge coupling is automatic in the
5-dimensional fermion action \diracf, if the original 4-dimensional Dirac
action \dirac\ is chiral.  In particular, one sees that if $q$ is, say,
left-handed,
then $\P_q$ is also left-hand, whereas $\Pb_q$ and $b_q$ are right-handed,
and oppositely for the anti-quark quartet.

        Let us find the form of the free propagators.  Because of the
presence of the
kernel there is a mixing of $q$ and $ba-q$.  The quadratic approximation of
the action is:
\eqn\fermions{\eqalign{ I_q= \int d^4 x dt
\ [ & b_q ^\dagger(   \partial_5  -  \partial_\m\partial_\m + m^2)  q
  - q^\dagger(  - \partial_5  -  \partial_\m\partial_\m + m^2)  b_q  \cr
    &  + b_q ^\dagger (\g_\m \p_\m + m) b_q
        + {\rm ghost \ terms}  ]
 }}
This gives the  following matrix propagator, between the independent pairs
  $(q^\dagger, b^\dagger_q)$ and $(q , b _q)$:
\eqn\prop{\eqalign{
 \pmatrix{ {{m+i\g_\m p_\m }\over {\omega^2 +(p^2 + m^2)^2}} &
{{1}\over {i\omega +p^2 + m^2}}
\cr
\ &\  \cr
 {{ 1 }\over {-i\omega +p^2 + m^2}}
 & 0
\cr  } }}
If we integrate the $q^\dagger-q$ matrix element over $\omega$, we
obtain the usual free Dirac propagator
\eqn\iprop{\eqalign{   { {1} \over  {\pi} } \int d \omega
{{m+i\g_\m p_\m }\over {\omega^2 +(p^2 + m^2)^2}}
 = {{m+i\g_\m p_\m }\over {p^2 + m^2}}.}}
There are closed loops of fields
$q$ and $q^\dagger$, but the associated ghosts $\Psi_q$ and $\bar\Psi_q$
have retarded propagators so  are no closed ghost
loops, just as in the bose case. The topological quantum
field theory in five dimensions operates for the spinors and their  topological
ghosts and Lagrange multipliers in the same way as it does for gauge fields.
We have a well-defined perturbation theory for the system of spinors and
gauge fields in five dimensions, and the familiar arguments about
renormalizability apply.

\appendix E{Anomalies}

 The five-dimensional theory is a local quantum quantum field theory
that is renormalizable by power counting. Our aim is that the
generating functional of Green functions must be constructed from the
requirement of $s$ and $\WW $ invariances. Then, observables are defined
from the operators that are the cohomology with ghost number zero of of
$\WW $, with expectation values taken at equal times.

The possibility of renormalizing the theory in five dimensions, while
 maintaining the two symmetries can be jeopardized by the presence of
 anomalies.  Such anomalies can  destroy the theory in two different
 ways, since the $s$-symmetry ensures the existence of a Fokker--Plank
 equation and the $\WW $-symmetry ensures that the drift forces give
 gauge invariant observables.  Our knowledge
of the four-dimensional theory indicates that some kind of breakdown
must occur in the five-dimensional approach when one couples a chiral
fermion to the theory, or a set of chiral fermions without couplings
that ensure anomaly compensations.

A primary check is thus to verify whether radiative corrections can
break the Ward identities of $s$ and $\WW $ symmetries. This can be done
by computing the possible obstructions of the Ward identity in five
dimensions and the values of their coefficient.  But, as already
mentioned, the currents and their conservation laws in four and five
dimensions are not related in an obvious way. For instance, if one
calls $J_\m(x)$ a conserved current of the four-dimensional theory,
the insertion of $\pa_\m J_\m(x,t)$ in correlation functions with
fields and/or operators taken at different values of the fifth
time $t$ has no reason to give zero.

Since we cannot rely on the above approach to understanding the effect
 of the anomaly, we are led to investigate the possibility that, when
 one computes insertions of four-dimensionally conserved currents in
 Green functions, their conservation is not fulfilled in the limit of
 equal fifth times for all arguments of the Green function.

The search of anomalies relies therefore on an indirect argument.
 Knowing from general principles that the symmetry of the theory in a
 given slice is the ordinary four-dimensional gauge symmetry, we can
 select and compute in five dimensions the possible anomalous vertices
 the four-dimensional theory.  We will do so, and find an ambiguity in
 the equal all fifth time limit for the external legs.  We will
 see that this ambiguity is rooted in the usual problem of computing
 Feynman diagrams with superficial linear divergences.

Before doing this important calculation, we will nevertheless show the
existence of a cocycle for the combined $\WW $ and $s$ symmetries.  Its
overall coefficient vanishes perturbatively (which indicates that
 no anomaly occurs in five dimensions, for all Green functions). However,
this cocycle has an interesting form, and  its role in quantum
field theory could be important. We  could not find its interpretation,
but it is
striking that it is related in a formal way to the usual
four-dimensional consistent anomaly by  descent
equations.

  The possible obstructions to the Ward identity for the $\WW $ and $s $
symmetries in five dimensions must be 5-forms, with ghost number (1,0)
and (0,1) respectively. We call them $\Delta^{(1,0)}_5$ and
$\Delta^{(0,1)}_5$. The ghost unification allows us to define
$\Delta^{1}_5=\Delta^{(1,0)}_5+\Delta^{(0,1)}_5$
Since
$(s+\WW )^2=0$,  $\Delta^{1}_5$ must satisfies the
 consistency equation for the $s$ and $\WW $ symmetries:
\def\ss{(s+\WW )}
 \eqn\wz{\eqalign{\ss\Delta^1_5
+d\Delta^2_4=0 }}
 If we assume that only exterior products play a role, we can easily
manipulate this equation. We use that
$\ss d \Delta^1_5=0$, and thus
 $~d  \Delta^1_5 +\ss \Delta^0_6=0~$. Thus $\ss d\Delta^0_6=0$. Since  $d
\Delta^0_6$ has ghost number 0, it is a 7-form that can only depend on
$A$ and
$dA$, and since the symmetry is $\ss A=\Psi -D(c+\l)$, the only
possibility is that $d \Delta^0_6=0$, which implies in turn that $
\Delta^0_6  $ is an invariant polynomial in the Yang--Mills curvature
$F=dA+AA$, that is, $ \Delta^0_6 =\Tr(FFF)=d Q_5(A, F)$, where $Q_5(A,
F)$ is a Chern class of rank 5. We thus have that the solution of the
Wess and Zumino consistency equation
\wz\ is obtained from the piece with ghost number 2 in the following
identity satisfied by
$\D_6(F + \P + \F) = \Tr[(F + \P + \F)^3]$:
\eqn\wzr{\eqalign{(s+\WW +d) \Delta_6(F+ \Psi+ \Phi)=0 }}
 We thus  have:
\eqn\wzr{\eqalign{  &\WW \Tr (  \Psi F F) =0\cr
 &\ss \Tr (  \Psi F F)+  d \ \Tr (\Psi\Psi F+FF\Phi)=0
}}
and thus
\eqn\wzsol{\eqalign{\Delta^{1,0}_5& = \Tr (  \Psi F F)\cr
\Delta^{0,1}_5& = 0  \quad   }}
 is a candidate
for   the anomaly.

On the other hand, due to the  Chern--Simons formula,
\eqn\cs{\eqalign{ \Tr[(F + \P + \F)^3]
= (d + s + \WW )Q_5(A+c+\lambda, F + \P + \F)
}}
we have
\eqn\wzrab{\eqalign{  \Tr (  \Psi F F)=  d \  \Tr
\Big[       c {{\delta }\over {\delta  A}}|_F
Q_5(A , F  )
+
\Psi {{\delta }\over {\delta  F}}|_A
Q_5(A , F  )
+ s Q_5(A , F  ) \Big]  }}
Therefore, in five dimensions, $\Tr ( \Psi F F)$ is locally the sum of
$d$- and $s$- exact terms. However, although it is $\WW $-invariant, it
not $\WW $-exact, up to $d$- and $s$-exact terms,
 and therefore we can identify $\Tr ( \Psi F F)$ as
part (and probably the unique element) of the cohomology with ghost
number one for the $\WW $-symmetry.  Due to this fact, we can probably
safely call $\Tr ( \Psi F F)$ the consistent five dimensional anomaly.

One recognizes among all  terms in the right hand side of
\wzrab\ the interesting piece:
$\pa_t \Delta_4^1$, where
\eqn\lb{\eqalign{  \Tr   \Delta_4^1=
\Tr   \Big[
 c {{\delta }\over {\delta A}}|_F Q_5(A , F ) \Big]  }}
is nothing else that
the consistent four-dimensional anomaly that can be directly related
to the ABBJ triangle anomaly once it is inserted on the right hand
side of the Ward identity in four dimensions.  All this suggests that adding
$Q_5(A, F)$ could become an interesting issue.

Although, it is certainly natural to interpret the existence of the
cocycle $\Tr (
\Psi F F)$ as the origin of the anomaly that must occur when the theory
 is coupled to  chiral four-dimensional spinors, we have not
been able yet to see the way $\Tr (
\Psi F F)$  plays a role in the five-dimensional theory.
 This does not mean however that the cocycle doesn't play
a role, for instance in Fujikawa type manipulations when on reduces the
theory in four dimensions.

Consider now practical computations, to understand how the anomaly
will manifest itself. We can take the case of a single spinor $q$, and
introduce the vector and axial currents $J^\m= \bar q \g ^\m q$ and
$J_5^\m= \bar q \g^5\g ^\m q$. The anomaly questions amounts to
compute the form factor of the 1PI vertex in four dimensions
$T^{\a\b}=<K_\m J_5^\m (K),J^\a (k), J^\b(k')>$ that is proportional
to $\e _ {\a\b\c\g} k^\a {k'}^\b$. Here we assume that the vector
current is conserved, and $K=k+k'$.

The free propagators of  spinors in five
dimensions are given by \prop.
 $T^{\a\b} (k,k',T,t,t') $ is thus given by
 the following 5-dimensional Feynman
integral
\eqn\triangle{\eqalign{\quad\quad\quad
\int &
d\Omega
d \omega
d \omega '
d^4p\
\exp i( \ \Omega T + \omega t  +\omega' t' \ ) \cr
&  \tr \big(K_\m\g^\m \g^5
{{(p-k)_\r \g^\r}\over{\omega^2 + ((p-k)_\r (p-k)^\r)^2}}
   \cr
& \g^\a
{{p_\r \g^\r}\over{{\Omega }^2 + (p_\r {p}^\r)^2}}
\g^\b
{{(p+k')_\r \g^\r}\over{{\omega '}^2 + ((p+k')_\r (p+k')\r)^2
}} \big) \cr
}}
We can use the translation invariance along the fifth time direction, and
set $T$ at the origin of time. It is thus sufficient to compute the
integral at $T=0$. We can  thus investigate carefully
the various way one can approach the value
$T=t=t'=0$ that must give the result of the four-dimensional theory.

We first observe, that if we set brutally $T=t=t'=0$ the integral
\triangle\ is just equal to the ordinary four-dimensional triangle
diagram. This is easily seen by performing the integration over
$d\Omega$, $d \omega $ and $d \omega$ by using Cauchy theorem for
picking out the poles in $\Omega$, $ \omega $ and $ \omega'$, which
gives:
\eqn\abbj{\eqalign{
 T^{\a\b} (k,k',T=t=t') =
\int
d\Omega
d \omega
d \omega'
d^4p \ &
\tr \big( K_\m\g^\m \g^5
{{(p-k)_\r \g^\r}\over{\omega^2 + ((p-k)_\r (p-k)^\r)^2}}
\g^\a
\cr&
{{p_\r \g^\r}\over{{\Omega }^2 + (p_\r {p}^\r)^2}}
\g^\b
{{(p+k')_\r \g^\r}\over{{\omega '}^2 + ((p+k')_\r (p+k')\r)^2}}  \big)
}}
When one extracts the term proportional to $
\e _{\a\b\m\n} k^\m{k'}^\n$, this  expression has the well known linear
divergence which provides a non-vanishing value for the anomaly.

If, on the other hand, we set $t$ and $t'$ different from zero, the
divergence in the integration over $p$ is regularized due to
the exponentials in $t$.  We can compute the integral explicitly for
small values of $t$ and $t'$, and one finds that the anomaly
coefficient is proportional to:
 \eqn\anomaly{\eqalign{
\e _{\a\b\m\n} k^\m{k'}^\n  {{|t-T|+|t'-T|}\over {|t-T|+|t'-T|+|t-t'|}} \ .
}}
This expression exhibits an ambiguity in the equal-time limit
$t=t'=T$. The ratio of absolute values varies between $1/2$ and
$1$.  It equals unity if one first sets $t=t'$ and then $t=T$; but it
equals $1/2$ if one first sets $t=T$ and then
$t'=T$.  This contrasts with the tree-level property that the correlation
functions in four dimensions are obtained as a smooth limit of their
counterparts in five dimensions.

The ambiguity of the expression \anomaly\ is related to the property that
first setting  $t=t'=T$ causes superficially linear divergences. We thus
foresee
that only the triangle  can lead to an ambiguity in the limit $t=t'=T$;
otherwise the fifth time acts an invariant regulator.

\appendix F{Absence of fermion doubling}

        We have seen in our analysis of anomalies that they do not appear
in the
5-dimensional theory per se, but rather when the fifth time is restricted to a
slice.  This suggests that the absence of fermion doubling in the 5-dimensional
formulation is practically automatic, and does not require the introduction of
domain walls \kaplan, \neubergerr, \shamir.

        Indeed, consider the matrix form of the free Dirac action
\fermions\ between
the independent pairs
$(q^\dagger, b^\dagger_q)$ and $(q , b _q)$,
\eqn\mdact{\eqalign{
 \pmatrix{ 0 &
\partial_5  +  \partial^2 - m^2
\cr
\ &\  \cr
  \partial_5  -  \partial^2 + m^2
 & \g_\m \p_\m + m
\cr  },}}
where $\p^2 \equiv \partial_\m\partial_\m$ for $\m = 1,...4$, is the
4-dimensional lattice Laplacian.  We shall use a standard lattice
discretization of the operators that appear here, $\p_5 \to (\p_5)_d$ etc.,
where the discretized operators are defined by
\eqn\discrete{\eqalign{
(\p_5)_d q(x,t) & \equiv \demi [q(x,t+1) - q(x,t-1)]  \cr
(\p^2)_d q(x,t) & \equiv \sum_{\m=1}^4
  [q(x+\hat{\m},t) + q(x-\hat{\m},t) - 2q(x,t)]
\cr (\p_\m)_d q(x,t) & \equiv \demi [q(x+\hat{\m},t) - q(x-\hat{\m},t)], }}
where $\hat{\m}$ is a unit vector in the $+\m$-direction.  We have
preserved hermiticity properties, so $(\p_5)_d$ and
$(\p_\m)_d$ are anti-symmetric whereas $(\p^2)_d$ is symmetric.
The trick is that we discretized the 4-dimensional lattice laplacian
$\p^2$ and the lattice derivatives $\p_\m$ independently, so
$(\p^2)_d \neq \sum_{\m=1}^4[(\p_\m)_d]^2$.  It will turn out
that we never have to invert the Dirac operator but only the even
chirality operators that appear in the off-diagonal matrix elements of the
matrix $\mdact$.  This simplifying property results from our choice of the
kernel, $K = m + \g_\m D_\m$, corresponding to $M = m$.

        All these operators are
diagonalized by lattice Fourier transformation, and the matrix \mdact\
becomes in terms of lattice momenta $\theta_\m$ and $\theta_5$,
\eqn\fmdact{\eqalign{
 \pmatrix{ 0 &
i \sin \theta_5   - Q_0
\cr
\ &\  \cr
  i \sin \theta_5  + Q_0
 & K_0
\cr  },}}
where $Q_0 \equiv m^2 + \sum_{\m = 1}^4 2(1 - \cos \theta_\m)$, and
$K_0 \equiv i \sin \theta_\m \g_\m + m$.

        The propagator is given by the inverse matrix,
\eqn\fmprop{\eqalign{
 \pmatrix{ { {K_0} \over {\sin^2 \theta_5   + Q_0^2} } &
{ {1} \over {i \sin \theta_5   + Q_0} }
\cr
\ &\  \cr
  { {1} \over {i \sin \theta_5   - Q_0} }
 & 0
\cr  }.}}
The free 4-dimensional lattice $q^\dagger-q$ propagator in momentum
space is obtained at equal fifth time, namely
\eqn\qprop{\eqalign{  S(\theta_\m) =
{ {1} \over {2\pi} }
\int_{-\pi}^{\pi} d \theta_5 { {K_0} \over {\sin^2 \theta_5   + Q_0^2} }
= { {K_0} \over {Q_0 (1   + Q_0^2)^{1/2}} }.}}
In the continuum limit, this integral gets contributions from the
neighborhood of $\theta_5 = 0$ and $\theta_5 = \pi$ which reflects fermion
doubling in $\theta_5$.  However there is no doubling of the
physical 4-dimensional propagator.  Indeed in the continuum limit,
$Q_0^2 = O(a^4)$ is negligible compared to
$1$, where $a$ is the lattice spacing, and we obtain
\eqn\aqprop{\eqalign{  S(\theta_\m) \approx
 { {K_0} \over {Q_0 } }
  = { {m + i \sum_{\m = 1}^4 \sin \theta_\m \g_\m}
   \over {m^2 + \sum_{\m = 1}^4 2(1 - \cos \theta_\m) } }.}}

        No fermion doubling occurs here.  As asserted, the Dirac
operator is never inverted, but only the operator
$i \sin \theta_5 + m^2 + \sum_{\m = 1}^4 2(1 - \cos \theta_\m)$ that has
even chirality.  The denominator of the last expression cannot be
factorized on the lattice, $Q_0 \neq K_0K_0^\dagger$, but it does factorize in
the continuum limit, when $\sin \theta_\m \to q_\m$, and
$2(1 - \cos \theta_\m) \to q_\m^2$.

        As regards practical numerical
simulation, a possible advantage of the lattice discretization described
here with respect to domain-wall fermions is that every hyperplane
$x_5 = {\rm const}$ may be used for 4-dimensional fermions, not just
the one domain wall \kaplan, \neubergerr, \shamir.


\footatend\vfill\supereject\immediate\closeout\rfile\writestoppt
\baselineskip=14pt\centerline{{\bf References}}\bigskip{\frenchspacing%
\parindent=20pt\escapechar=` \input refs.tmp\vfill\eject}\nonfrenchspacing

\bye